\documentclass[aps,
prd,
notitlepage,twocolumn,showpacs,superscriptaddress,nofootinbib,preprintnumbers]{revtex4-2}  
\usepackage{graphicx}  
\usepackage{dcolumn}   
\usepackage{bm}        
\usepackage{amssymb}   
\usepackage{amsmath}
\usepackage{physics}
\usepackage{slashed}
\usepackage{mathtools}
\PassOptionsToPackage{dvipsnames}{xcolor}
\PassOptionsToPackage{table}{xcolor}
\usepackage[colorlinks=true,citecolor=teal,urlcolor=teal,linkcolor=Maroon]{hyperref}
\usepackage{cleveref}
\usepackage{orcidlink}
\usepackage{tikz}
\usepackage{subcaption}
\usepackage{diagbox}
\usetikzlibrary{decorations.pathmorphing}
\usetikzlibrary{decorations.markings}
\usetikzlibrary{positioning, shapes, snakes, arrows}
\usetikzlibrary{patterns.meta}
\usetikzlibrary{arrows.meta}

\usepackage{soul}
\usepackage{ragged2e}
\usepackage[compatibility=false]{caption}

\usepackage{tikz}
\usetikzlibrary{decorations.pathmorphing}
\usetikzlibrary{decorations.markings}
\usetikzlibrary{positioning, shapes, decorations, arrows}
\usetikzlibrary{patterns}

\tikzset{
  externalShorter/.style={
    shorten <=2mm,
    shorten >=2mm
  },
  wl/.style={line width=1pt},
  graviton/.style={line width=.8pt, -latex,decorate, decoration={snake, segment length=4pt,amplitude=1.8pt, pre length=.15cm, post length=.25cm}},
    worldlineStatic/.style={dotted, line width=1pt},
	worldline/.style={gray, line width=1pt},
	worldlineBold/.style={black, line width=.6pt},
	zUndirected/.style={line width=1pt},
	zParticle/.style={line width=1pt,postaction={decorate},decoration={markings,mark=at position .6 with {\arrow[#1]{latex}}}},
	zParticleTest/.style={white,line width=1pt,postaction={decorate},decoration={markings,mark=at position .6 with {\arrow[#1]{latex}}}},
  example1/.style={
    draw=none,
    postaction={decorate},
    decoration={
      markings,
      mark=at position #1 with {
        \arrow[
          black,
          scale=2.4
        ]{Straight Barb}
      }
    }
  },
  example2/.style={
    white,
    line width=1pt,
    postaction={decorate},
    decoration={
      markings,
      mark=at position .6 with {
        \arrow[
          draw=black,
          scale=1.5,
          #1
          ]{latex}
      }
    }
  },
	zParticle2/.style={line width=1pt,postaction={decorate},decoration={markings,mark=at position .75 with {\arrow[#1]{latex}}}},
	zParticle1/.style={line width=1pt},
	worldlineCut/.style={dotted,line width=1pt,postaction={decorate},decoration={markings,mark=at position .7 with {\arrow[#1]{latex}}}},
	worldlineCut2/.style={dotted,line width=1pt,postaction={decorate},decoration={markings,mark=at position .6 with {\arrow[#1]{latex}}}},
	zParticleF/.style={line width=1pt,postaction={decorate}},
	cscalar/.style={line width=1pt,postaction={decorate},decoration={markings,mark=at position .6 with {\arrow[#1]{latex}}}},
	cscalar2/.style={line width=1pt,postaction={decorate},decoration={markings,mark=at position .8 with {\arrow[#1]{latex}}}},
	photon/.style={line width =.8pt, decorate, decoration={snake, segment length=4pt, amplitude=1.8pt,  pre length=.1cm, post length=.1cm}},
	photonTest/.style={
    line width =.8pt,
    decorate,
    decoration={
      snake,
      segment length=5pt,
      amplitude=1.4pt,
      pre length=.0cm,
      post length=.0cm}},
	photonRed/.style={red, line width =.8pt, decorate, decoration={snake, segment length=4pt, amplitude=1.8pt,  pre length=.1cm, post length=.1cm}},
	cross/.style={cross out, line width =.8pt, draw=black, minimum size=2*(#1-\pgflinewidth), inner sep=0pt, outer sep=0pt},
cross/.default={4pt},
  dottedLine/.style={
    dotted, thick
  }
}
\usetikzlibrary{calc}

\pgfdeclarelayer{foreground}
\pgfsetlayers{main,foreground}

\newcommand{\Lagr}{\mathcal{L}}

\newcommand{\cO}{\mathcal{O}}
\newcommand{\cM}{\mathcal{M}}

\newcommand{\cY}{\mathcal{Y}}

\newcommand{\mn}{{\mu\nu}}
\newcommand{\ab}{{\alpha\beta}}

 \newcommand{\Ysw}[3]{{}_{#1}Y_{#2 #3}}
 \newcommand{\Ssw}[3]{{}_{#1}S_{#2 #3}}
 \newcommand{\Dwig}[3]{D^{#1}_{#2 #3}}

\newcommand{\iO}{i 0^{+}}
\newcommand{\ePM}{\epsilon_{\rm PM}}

\newcommand{\zenodo}{\href{https://doi.org/10.5281/zenodo.24303071}{\tt zenodo.org} \cite{zenodo}}

\begin{document}

\title{A Kerr-black-hole worldline action to $\cO(G^5)$}

\author{Yilber Fabian Bautista\,\orcidlink{0000-0001-6255-5675}}
\affiliation{Higgs Centre for Theoretical Physics, School of Physics and Astronomy, The University of Edinburgh, \\
Edinburgh EH9 3JZ, Scotland, UK}
\author{Mathias Driesse\,\orcidlink{0000-0002-3983-5852}}
\affiliation{Institut f\"ur Physik, Humboldt-Universit\"at zu Berlin,
 10099 Berlin, Germany}
 \author{Kays Haddad\,\orcidlink{0000-0002-1182-2750}}
\affiliation{Institut f\"ur Physik, Humboldt-Universit\"at zu Berlin,
 10099 Berlin, Germany}
\author{Gustav Uhre Jakobsen\,\orcidlink{0000-0001-9743-0442}}
\affiliation{Institut f\"ur Physik, Humboldt-Universit\"at zu Berlin,
 10099 Berlin, Germany}

\date{\today}

\begin{abstract}
We establish a worldline action which correctly reproduces Kerr-black-hole dynamics up to fifth order in Newton's constant in a physical post-Minkowskian counting.
This encompasses operators up to second order in the Riemann curvature tensor and up to fourth order in the spin tensor.
Doing so entails performing the full effective-field-theory computation of the amplitude for a gravitational-wave scattering off of a spinning compact object up to four-loop order, and matching the result to the predictions from linear black hole perturbation theory.
Quadratic-in-curvature operators which must be added to match successfully include time-reversal anti-symmetric ones at linear and cubic spin orders, capturing absorption across the black hole horizon, and conservative quadratic-in-spin ones.
As an immediate application of our action, we compute the leading-order change in spin magnitude, and reproduce the leading change in mass, for a Kerr black hole in a scattering event with a second
compact body, considering a generic orientation of the black hole's spin. 

\end{abstract}


\preprint{HU-EP-26/35}

\maketitle


\tableofcontents

\section{Introduction}

Gravitating compact objects, such as black holes (BHs) and neutron stars (NSs), appear indistinguishable from elementary point particles when observed from sufficiently far away.
This behaviour originates from a separation of scales, made precise as follows.
Let $r$ denote the characteristic size of the object.
If it possesses rotational angular momentum $J$, its spatial extent is supplemented by the additional length scale $\mathcal{J}\equiv GJ/r$, with $G$ Newton's constant.
Probing the object with radiation of wavelength $\lambda$ does not reveal its spatial scales to a distant observer so long as $r/\lambda \ll 1$ and $\mathcal{J}/\lambda\ll1$: the object effectively appears point-like, with its dynamics described by a worldline minimally coupled to gravity.

As $\lambda$ decreases, the object's extended nature eventually becomes apparent to the distant observer.
An accurate description of its dynamics then requires the inclusion of additional effective operators supplementing  the point-particle Lagrangian.
Beyond parametrizing deviations from point-like behavior, the coefficients of these effective operators distinguish the identity of the compact object.
In the absence of rotation, BHs and NSs differ in their tidal responses to external gravitational fields,  an $\cO((r/\lambda)^5)$ effect \cite{Mora:2003wt,Goldberger:2004jt}.
On the other hand, these objects have distinguishable spin-induced quadrupoles when they rotate, meaning their effective descriptions would differ already at $\cO((\mathcal{J}/\lambda)^2)$ \cite{Thorne:1980ru,Laarakkers:1997hb,Porto:2005ac}.
A natural question, then, is, given a compact object, what low-energy effective action describes that object's dynamics?

In this work, we seek to answer this question in the case of rotating, i.e. Kerr, BHs, up to the order where their tidal response may first enter.
Denoting its mass by $m$, the characteristic size of a Kerr BH is proportional to its Schwarzschild radius $r=r_{s}=2Gm$, meaning our targeted precision is commensurate with the fifth order in a post-Minkowskian (PM) expansion, $\cO(G^5)$.
Cosmic censorship folds the spin multipole expansion into this PM target; namely, for the BH's singularity to lie behind an event horizon, its rotational angular momentum must satisfy $J/Gm^2\leq 1$.
Our goal thus becomes to identify the action capturing all effects of the order
\[
  \left(\frac{r_{s}}{\lambda}\right)^{n+1}\left(\frac{\mathcal{J}}{\lambda}\right)^{s}
\]
for $n+s\leq4$.
For $n=0$ and any $s$, appropriate effective worldline descriptions are those in refs.~\cite{Levi:2015msa,Vines:2017hyw}.

In standard effective-field-theory (EFT) fashion, we set off with an ansatz for our action that reaches the precision we seek.
As we will argue below, this includes operators up to quadratic order in curvature and quartic order in spin.
Then, drawing the connection to the description of BHs in general relativity (GR) means performing a matching calculation: the same observable is computed in both frameworks, and any free parameters in the EFT description are fixed by demanding equivalence.
The amplitude for the scattering of a gravitational wave off of a spinning BH is suitable in this regard.
However, as has been advocated in refs.~\cite{Ivanov:2024sds,Ivanov:2026icp,Bautista:2026qse}, its surrogates in the form of the scattering phase shift and absorption probability are more convenient to match.\footnote{In the below, we use the term ``observable'' to refer to the amplitude, scattering phase shift, and absorption probability when the distinction is not important.}
The GR prediction for these latter quantities is extracted from linear BH perturbation theory (BHPT) \cite{Teukolsky:1973ha,Press:1973zz,Teukolsky:1974yv,Matzner:1977,Matzner:1978,Mano:1996vt,Mano:1996mf,Dolan:2008kf,BHPToolkit} 

On the EFT side, we will employ the framework of worldline quantum field theory (WQFT) \cite{Mogull:2020sak,Jakobsen:2021zvh,Jakobsen:2022psy,Haddad:2024ebn}.
A successor to earlier effective worldline descriptions of gravitating bodies \cite{Goldberger:2004jt,Porto:2005ac,Levi:2015msa,PORTO20161}, WQFT efficiently adapts QFT techniques to a classical setting, with a particular focus on PM dynamics; see also ref.~\cite{Kalin:2020mvi,Liu:2021zxr}.
Recently, it has also been a prominent tool applied to this specific problem of mapping effective descriptions of BHs to BHPT in the absence of spin \cite{Bjerrum-Bohr:2025bqg,Bautista:2026qse,Bjerrum-Bohr:2026fhs,Bautista:2026fcp,Brunello:2026rdk}, with the state of the art using this method reaching $\cO(G^5)$ \cite{Brunello:2026lzf}.
We build on this here, producing the first loop-level matching to BHPT involving spin, while also reproducing the recent spinless $\cO(G^5)$ result just cited.

This program of matching effective BH descriptions to GR is not new, but has seen rapid acceleration over the past several years. 
Roused by the question of the tree-level Compton amplitude suitable for describing Kerr BHs, comparisons to GR were initiated within an entirely on-shell framework \cite{Guevara:2018wpp,Chung:2018kqs,Guevara:2019fsj,Arkani-Hamed:2019ymq,Bautista:2021wfy,Bautista:2022wjf,Cangemi:2022bew,Aoude:2023fdm,Aoude:2023dui,Bautista:2023sdf,Cangemi:2023bpe,Bautista:2024emt,Bjerrum-Bohr:2026low}.
Many comparisons centering effective actions other than WQFT also exist \cite{Bern:2020buy,Charalambous:2021kcz,Scheopner:2023rzp,Ivanov:2024sds,Caron-Huot:2025tlq,Ben-Shahar:2025tiz,Combaluzier--Szteinsznaider:2025eoc,Ivanov:2026icp}, with the distinct advantage of involving Wilson coefficients which are easier to interpret physically than those in an on-shell amplitude. 
Other remarkable innovations include the Born series and  effective wave equation method of refs.~\cite{Correia:2024jgr,Caron-Huot:2025tlq} and the shell EFT of ref.~\cite{Kosmopoulos:2025rfj}, each of which have culminated in results for Schwarzschild BH Love numbers to unprecedented PM orders \cite{Solon:2026ubm,Correia:2026utp,Combaluzier--Szteinsznaider:2025eoc}.
We also note interesting recent progress in the matching of worldline actions to stellar perturbation theory, effectively conceiving of NSs as deviations from BHs \cite{Saketh:2026trm,Saketh:2026bcx}.

In this work we do not yet answer the question which initiated some of this flurry of activity, namely the effective description of Kerr BHs beyond quartic order in spin.
However, our work here brings us right to this precipice: extending the matching in this paper to the next PM order will pin down this effective description.

Underlying much of this interest is the goal of describing binary-BH dynamics to high accuracy, naturally hinging upon precise modelling of a single BH.
In this context, our results here are well timed.
We discover the need for operators simultaneously quadratic in spin and curvature in order to match BHPT.
Relevant to binary-BH scattering at $\cO(G^6)$, these operators are indispensable for augmenting the three-loop linear-in-spin results of ref.~\cite{Jakobsen:2023hig} to the next spin order, or the quadratic-in-spin waveform of ref.~\cite{Bohnenblust:2025gir} to the next loop order.
Our action also has relevance to post-Newtonian (PN) approaches.
Specifically, those of refs.~\cite{Levi:2020uwu,Kim:2022bwv,Kim:2021rfj} incorporate similar quadratic-in-spin and curvature operators, important from 5PN order, whose coefficients can now be fixed by reference to our results here.

Setting aside involved high-loop endeavors, the matching we perform here finds immediate application to mass- and spin-changing effects in two-body scattering.
Our action includes a sector which accommodates time-asymmetric effects.
Upon matching to BHPT, we identify this sector as corresponding to absorption of gravitational radiation by the BH.
We are then easily able to quantify the leading---i.e. one-loop---changes to the mass and spin magnitude of a Kerr BH due to absorption in binary scattering~\cite{Poisson:2004cw,Taylor:2008xy,Goldberger:2020wbx,Porto:2007qi,Goldberger:2020fot,Saketh:2022xjb,Jones:2023ugm,Goldberger:2005cd}. In particular, 
we reproduce the change in mass result reported in eq. $(83)$ in ref.~\cite{Bautista:2024emt} for generic spin orientations, and calculate for the first time the leading-order change in spin for Kerr BHs.
We further validate our computational method by reproducing these same quantities for a Schwarzschild BH scattering off a secondary point-like object, reported for the first time in refs.~\cite{Goldberger:2020wbx} and~\cite{Cipriani:2026myb}, respectively.

This paper is organized as follows.
In \Cref{sec:Action} we present the necessary ideas for constructing an action relevant to Kerr BH dynamics up to $\cO(G^{5})$.
Matching to BHPT involves more considerations than in the absence of spin, which we lay out in \Cref{sec:Matching}.
Details needed to augment previous WQFT calculations to include spin and to go to four-loop order are given in \Cref{sec:Computation}.
Finally, the results of our matching to BHPT, as well as absorptive observables, are presented in \Cref{sec:Results}.
In \Cref{app:ChiConst,app:RecoilProof} we prove some convenient properties of our perturbative calculation, particularly arguing for the absence of recoil in our frame even in the presence of spin.
\Cref{app:absComputations} presents details of the computation of absorbed mass and spin magnitude.
Finally, \Cref{app:BHPTResults} collects the BHPT results which we match.

\section{The Kerr-black-hole action}\label{sec:Action}

We refer the reader primarily concerned with the result to \cref{eq:GenAction,eq:NonMinSplit,eq:KerrLinearR,eq:ConservativeR2,eq:AbsorptiveR2}.

We are interested in producing an effective model of a spinning BH in Einstein gravity, where the gravitational (bulk) degrees of freedom are governed by the Einstein-Hilbert action,
\begin{align}\label{eq:EH_action}
  S_{\rm EH}=-\frac{2}{\kappa^{2}_{d}}\int{\rm d}^{d}x\,\sqrt{-g}R,
\end{align}
comprised of the Ricci scalar $R$, the metric determinant $g$, and the $(d=4-2\varepsilon)$-dimensional coupling constant $\kappa_{d}^{2}=32\pi G\tilde\mu^{2\varepsilon}$, with $\tilde\mu$ a regularization scale.

In the bosonic-oscillator encoding of spin on the worldline of mass $m$, the oscillator $\alpha^{a}$ and its complex conjugate $\bar{\alpha}^{a}$ compose the spin tensor according to \cite{Haddad:2024ebn,Haddad:2025cmw}
\begin{align}\label{eq:SpinTensor}
  S^{ab}(\tau)=-2im\bar{\alpha}^{[a}(\tau)\alpha^{b]}(\tau),
\end{align}
where $X^{[a} Y^{b]} = \frac{1}{2} (X^a Y^b - X^b Y^a)$.
Latin indices from the beginning of the alphabet exist in the locally flat frame accessed by the vielbeins:
\begin{align}\label{eq:Vielbein}
  e^{\mu}_{a}(x)e^{\nu}_{b}(x)g_{\mu\nu}(x)=\eta_{ab}.
\end{align}
When oscillators (or the spin tensor) are written with Greek---i.e. curved-space---indices, we are abbreviating e.g. $\alpha^{\mu}=e^{\mu}_{a}\alpha^{a}$.

In this formulation, a spinning, gravitating compact object can be described by the action
\begin{equation}\label{eq:GenAction}
\begin{aligned}
  S_{\rm wl}=\int{\rm d}\tau&\left[-\frac{m}{2}g_{\mu\nu}(x)\dot{x}^{\mu}(\tau)\dot{x}^{\nu}(\tau)\right. \\
  &\left.-im\bar{\alpha}_{a}(\tau)\frac{\nabla\alpha^{a}(\tau)}{{\rm d}\tau}+\mathcal{L}_{\rm nm}\right].
\end{aligned}
\end{equation}
The first and second terms here, describing respectively the center-of-mass (i.e. spin-monopole) and spin-dipole dynamics, are universal for any gravitating object due to diffeomorphism invariance of GR.
The second term parallel-transports a bosonic oscillator along the worldline trajectory,
\begin{align}
  \frac{\nabla\alpha^{a}(\tau)}{{\rm d}\tau}=\dot{\alpha}^{a}+\dot{x}^{\mu}{{\omega_{\mu}}^{a}}_{b}\alpha^{b},
\end{align}
involving the spin connection ${\omega_{\mu}}^{ab}=e_{\nu}^{a}\partial_{\mu}e^{\nu b}+e_{\nu}^{a}e^{\sigma b}\Gamma^{\nu}_{\sigma\mu}$, defined in terms of the Christoffel symbol $\Gamma^{\mu}_{\alpha\beta}$.
Identifying the body requires knowledge of the non-minimal $\mathcal{L}_{\rm nm}$, which may enter already at $\cO(G^{5/2})$.

Perturbative computations require the expansion of the dynamical degrees of freedom in the action around a chosen background. 
Just as the trajectory is perturbed around a straight line background, so too are the bosonic oscillators expanded around asymptotic values; in all, the three quantities perturbed around their flat space values are\footnote{Perturbations of $\bar\alpha^{a}$ are complex conjugates of those of $\alpha^{a}$.}
\begin{align}
  x^{\mu}(\tau)&=v^{\mu}\tau+\delta x^{\mu}(\tau),\label{eq:ExpandWL} \\
  \alpha^{a}(\tau)&=\alpha^{a}_{0}+\delta \alpha^{a}(\tau), \\
  g_{\mu\nu}(x)&=\eta_{\mu\nu}+\kappa_{d}h_{\mu\nu}(x),\label{eq:ExpandMetric}
\end{align}
and the $\delta x^{\mu}$, $\delta\alpha^a$, and $h_{\mu\nu}$ are the dynamical degrees of freedom in our formalism.
From the background oscillators $\alpha^{a}_{0}$, one defines the background spin tensor, which satisfies the covariant spin-supplementary condition (SSC) $v_\mu S^{\mu\nu}_{0}=0$.

Determining the $\mathcal{L}_{\rm nm}$ which describes Kerr BHs up to $\cO(G^5)$ is the main goal of this work.
Realizing this objective first of all demands a classification of the operators which may enter up to this level of precision.

\subsection{Power counting}

As is standard in EFTs, dimensional analysis enables an estimation of the perturbative significance of allowed operators.
Since Newton's constant, in which we are perturbing, is dimensionful, such an analysis requires that we distinguish between units of length and mass.

Aside from the spin-dipole term, bosonic oscillators only appear in $\mathcal{L}_{\rm nm}$ in the guise of the spin tensor defined in \cref{eq:SpinTensor}, such that the constituents of the non-minimal Lagrangian carry dimensions
\begin{align*}
  [m]=[M],\quad [\dot{x}^{\mu}]&=1,\quad [S^{ab}]=[L][M],\\
  [G]=\frac{[L]}{[M]},\quad [R_{\mu\nu\alpha\beta}]&=[L]^{-2},\quad [\nabla^\mu]=[L]^{-1},
\end{align*}
with $[L]$ denoting dimensions of length, $[M]$ those of mass, and $\nabla^{\mu}$ the covariant derivative.
Consider then an operator in the action taking the schematic form
\begin{align*}
  \bar{C}\, S^j\nabla^k R^l,
\end{align*}
with Wilson coefficient $\bar{C}$.
To ensure that such an operator carries the same dimensions as the Lagrangian,
\begin{align*}
[\mathcal{L}_{\rm nm}]=[M],
\end{align*}
its coefficient must satisfy
\begin{align}\label{eq:CoeffPowerCounting}
  \bar{C}=m^{1-j}(mG)^{k+2l-j}C,
\end{align}
for some dimensionless $C$.

Now, we anticipate the identification of the (magnitude of the) spin tensor $S^\mn$ with the Kerr BH rotation parameter.
Thus, while every power of the spin tensor lessens the operator's explicit factors of $G$, the cosmic censorship bound on a BH's spin
\begin{align}\label{eq:ChiParameter}
  \chi^{2}=\frac{S_{\mu\nu}S^{\mu\nu}}{2G^{2}m^{4}}\leq1\,,
\end{align}
means each power implicitly contributes one order to the PM expansion.
Also, \cref{eq:ExpandMetric} means that $R\sim G^{1/2}$.\footnote{Technically, $h_{\mu\nu}\sim G^{1/2}$ as well so that actually $R\sim G$. However, in the extraction of Feynman rules the action is differentiated by the graviton field $h_{\mu\nu}$, so that, at leading order, each curvature contributes $G^{1/2}$ to vertices.}
Consequently, the PM relevance of our schematic operator is
\begin{align}\label{eq:OpPowerCounting}
  \bar{C}\,S^{j}\nabla^{k}R^{l}\sim\cO(G^{k+5l/2}),
\end{align}
independently of its spin order.

\Cref{eq:CoeffPowerCounting} introduces a dimensionless version of a generic Wilson coefficient.
Generally speaking, this coefficient is not just a number, but is rather a function
\begin{align}\label{eq:WCFunction}
  C=C(\chi)
\end{align}
of the dimensionless constant $\chi$ in \cref{eq:ChiParameter}.\footnote{We prove in \Cref{app:ChiConst} that this is indeed a constant despite its dependence on the dynamical spin tensor.}
Up to the precision in $G$ considered in this work, every Wilson coefficient we introduce is simply a polynomial in $\chi$, such that we may write
\begin{align}
  C(\chi)=c_0 + c_2 \chi^2 + c_4 \chi^4 +\dots.
\end{align}
This convenient decomposition is not a perturbative spin expansion, and is no longer admissible from $\cO(G^6)$ \cite{Bautista:2022wjf,Bautista:2024agp,Munna:2023wce,Goldberger:2020fot,Saketh:2022xjb}.

Going back to \cref{eq:OpPowerCounting}, operators relevant to our $\cO(G^{\leq5})$ target are either linear ($l=1$) or quadratic ($l=2$) in the Riemann curvature.
Let us then split our non-minimal Lagrangian as
\begin{align}\label{eq:NonMinSplit}
  \mathcal{L}_{\rm nm}=\mathcal{L}_{R}+\mathcal{L}_{R^{2}}+\dots,
\end{align}
and study each term individually.
The ellipsis indicates higher-curvature operators which are not relevant at our precision of interest.

\subsection{Linear order in curvature}

At linear order in curvature, the number of covariant derivatives---which act only on curvature tensors---in an operator is tied to the number of spin tensors.
This is because Lorentz contracting a covariant derivative  with the curvature tensor, worldline four-velocity, or another covariant derivative produces an operator redundant with a higher-curvature one.
The upshot is that a linear-in-curvature operator with $k$ covariant derivatives must include $k+2$ factors of the spin tensor which absorb the indices of the derivatives.\footnote{In more than four dimensions evanescent operators with more than $k+2$ factors of the spin tensor are allowed. These, however, contribute redundantly to four-dimensional observables.}

For the precision targeted in this work, it then follows from \cref{eq:OpPowerCounting} that we should construct operators up to fourth order in the spin tensor.
The most general set of operators in the bosonic-oscillator framework was written down in refs.~\cite{Haddad:2024ebn,Haddad:2025cmw}.\footnote{Preceding this framework was the body-fixed-frame approach for the modelling of spin effects \cite{Porto:2005ac,Porto:2007qi}, where the most general set of operators at linear order in curvature was presented in ref.~\cite{Levi:2015msa}.
See also refs.~\cite{Goldberger:2017ogt,Li:2018qap,Liu:2021zxr,Ben-Shahar:2023djm,Scheopner:2026czu,Scheopner:2026sok} for yet more constructions.}
In ref.~\cite{Haddad:2025cmw}, considerations from refs.~\cite{Levi:2015msa,Vines:2017hyw} were used to identify the Kerr-relevant Wilson coefficients of this set of operators.
Specifically, the Kerr values are those which produce an exponential dependence of the stress-energy tensor (the single-graviton emission amplitude) on the spin vector.

Seeking to avoid ambiguities in dimensional regularization, we recast here the action of ref.~\cite{Haddad:2025cmw} without relying on the four-dimensional Levi-Civita tensor, writing 
\begin{subequations}\label{eq:KerrLinearR}
\begin{align}
  \mathcal{L}_{R}^{\rm Kerr}=\sum_{n=2}^{4}\mathcal{L}_{RS^n}^{\rm Kerr}+\cO(G^{11/2}),
\end{align}
where
\begin{align}
  \mathcal{L}_{RS^{2}}^{\rm Kerr}&=\frac{1}{8m}R_{\mu\nu\alpha\beta}S^{\mu\nu}S^{\alpha\beta}\label{eq:Spin2R1Ops} \\
  \mathcal{L}_{RS^{3}}^{\rm Kerr}&=\frac{1}{12m^{2}|\dot{x}|}S^{\mu\rho}\nabla_{\sigma}R_{\mu\rho\nu\lambda}\label{eq:Spin3R1Ops} \\
  &\times\left[\dot{x}^{\lambda}(S\cdot S)^{\sigma\nu}+S^{\nu\lambda}Z^{\sigma}\right]\notag \\
  \mathcal{L}_{RS^{4}}^{\rm Kerr}&=\frac{1}{48m^3}(S\cdot S)^{\rho\sigma}\nabla_{\rho}\nabla_{\sigma}R_{\mu\nu\lambda\beta}\label{eq:Spin4R1Ops} \\
  &\times\left[S^{\mu\nu}S^{\lambda\beta}-\frac{2\dot{x}^{\nu}\dot{x}^{\beta}}{\dot{x}^{2}}(S\cdot S)^{\mu\lambda}\right]\notag,
\end{align}
\end{subequations}
and $\dot{x}^{2}=g_{\mu\nu}\dot{x}^{\mu}\dot{x}^{\nu}$.
Spin-exponentiation of the single-graviton-emission amplitude determines only the coefficients of operators free of the SSC vector
\begin{align}
  Z^\mu=S^{\mu\nu}\dot{x}_{\nu}\,;
\end{align}
the rest are fixed through conservation of the covariant SSC
\begin{align}
  Z^\mu\overset{\tau\rightarrow\pm\infty}{\longrightarrow}0,
\end{align}
following the method in Appendix B of ref.~\cite{Haddad:2025cmw} to avoid introducing a second massive body.

\subsection{Quadratic order in curvature}\label{sec:R2Ops}

At this order, \cref{eq:OpPowerCounting} tells us we should consider operators free of covariant derivatives.
Now the presence of two curvature tensors introduces enough Lorentz indices to decouple the number of spin tensors and derivatives.
As \cref{eq:OpPowerCounting} is spin-independent, our precision target technically does not constrain the spin order we should consider;
we restrict ourselves to fourth order to remain consistent with the lower-curvature operators. 

Going beyond the power counting, the $\cO(R^2)$ operators that we will consider in the following have a concrete physical origin.
As elaborated in, e.g., refs.~\cite{Porto:2007qi,Goldberger:2020fot,Saketh:2022xjb,Saketh:2023bul,Ivanov:2024sds,Saketh:2024juq,Ivanov:2026icp}, a BH's multipolar tidal response to a perturbing field is represented by a dynamical degree of freedom on the worldline coupled to the external field.
Integrating out these multipoles in an adiabatic limit expands them in powers of the perturbing field and its derivatives.
At leading order in this adiabatic expansion, the resulting operators are simply the ones we present here.

This perspective is particularly important for understanding time-asymmetric effects, i.e. absorption, which arise in the BHPT result starting at $\cO(G^5)$ \cite{Bautista:2024emt,Goldberger:2020wbx,Porto:2007qi,Goldberger:2020fot,Saketh:2022xjb}.
Causal information encoded in the retarded two-point function of the integrated-out multipoles must be preserved in this adiabatic expansion.
This cannot be achieved in the time-symmetric (in-out) formalism employed so far.
We therefore further split the quadratic-in-curvature part of the action into
\begin{align}
  \mathcal{L}_{R^2}=\mathcal{L}_{R^2,{\rm cons}}+\mathcal{L}_{R^2,{\rm abs}}\,,
\end{align}
where the first term on the right-hand side encodes conservative, and the second absorptive, effects, treating each individually in what follows.

\subsubsection*{Conservative sector}

Conservative dynamics are time-symmetric, so we persist with the in-out formulation used thus far.
In this sector, ten operators with two curvature tensors may be enumerated up to fourth order in the spin tensor; see refs.~\cite{Haddad:2024ebn,Hoogeveen:2025tew}.
Rather than listing them all, we present only those needed to match a Kerr BH's response to an external gravitational perturbation:
\begin{equation}\label{eq:ConservativeR2}
\begin{aligned}
  &\mathcal{L}_{R^{2},{\rm cons}}=mG^{2}S^{\mu\nu}S^{\alpha\beta}\left[c_{(S^2B^2;0)}B_{\rho\mu\alpha}{B^{\rho}}_{\nu\beta}\right. \\
  &\left.+c_{(S^2RE;0)}R_{\mu\nu\alpha\rho}{E^{\rho}}_{\beta}+c_{(S^2E^2;0)}E_{\mu\alpha}E_{\nu\beta}\right] \\
  &+G^{4}m^{5}\chi^{2}c_{(S^0B^2;2)}B_{\mu\nu\rho}B^{\mu\nu\rho} \\
  &+G^2m\chi^2 c^{\prime}_{(S^2B^2;2)}B_{\mu\nu\rho}{B^{\mu\nu}}_{\alpha}(S\cdot S)^{\rho\alpha} \\
  &+G^4m^5\chi^4 c_{(S^0R^2;4)}R_{\mu\nu\rho\beta}R^{\mu\nu\rho\beta}.
\end{aligned}
\end{equation}
Spin powers in the coefficient subscripts refer to how many of the spin tensors introduce angular dependence to the operator, while the number after the semicolon counts factors of $\chi$.
Curvature tensors have been abbreviated as
\begin{align}
  E_\mn 
  \equiv  
  R_{\mu\alpha\nu\beta}\frac{\dot x^\alpha \dot x^\beta}{\dot x^2},\quad  B_{\mn\alpha}
  \equiv
  R_{\mn\ab}\frac{\dot x^\beta}{|\dot x|}
  \ .
\end{align}
In four spacetime dimensions, these two are sufficient for building a basis of operators, but in general dimensions the full Riemann curvature tensor is retained as well.

In both \cref{eq:KerrLinearR,eq:ConservativeR2} we have written operators using all components of the curvature and its covariant derivatives.
This stands in contrast to the common writing of non-minimal operators using projections of curvatures and their derivatives transforming in irreducible representations of $SO(d-1)$~\cite{Damour:1990pi,Damour:2009wj,Bini:2012gu,Ivanov:2026icp}.
In the latter writing, all indices should be projected onto the spatial subspace, i.e. by projecting out the $\dot{x}^{\mu}$ directions, with the remaining spatial indices made traceless and symmetrized.
We have opted not to work in an $SO(d-1)$ basis here such that we may more compactly write an explicit form of the action.

This comes at the cost, however, of needing the last two lines of \cref{eq:ConservativeR2} to match to BHPT.
In particular, for the Kerr values we determine in \cref{eq:BHPTMatchingCoeffs}, using an $SO(d-1)$ basis for \cref{eq:Spin4R1Ops} would lead to the cancellation of these two lines.
For example, the four-dimensional linear-in-curvature action of ref.~\cite{Haddad:2025cmw} produces the BHPT-matching tree-level amplitude at fourth order in spin.

In any case, the first three lines of \cref{eq:ConservativeR2} cannot be encapsulated in a linear-in-curvature basis regardless of the writing of \cref{eq:KerrLinearR}.
This follows from their explicit factors of $G$ being more than their factors of $\chi$, which cannot arise from an $\cO(R)$ operator that contributes on shell.\footnote{A linear-in-curvature operator including $\dot{x}\cdot\nabla$ would have a $G$-dependent Wilson coefficient, but such an operator vanishes on shell for single-graviton emission due to energy conservation.}
Though they resemble putative static tidal operators, Kerr BHs are known to be rigid under static disturbances \cite{Chia:2020yla,Charalambous:2021mea,Charalambous:2021kcz}.
We leave the study of their qualitative interpretation to future work, but note that similar operators with $G$-dependent coefficients have appeared also in refs.~\cite{Levi:2020uwu,Kim:2021rfj,Kim:2022bwv}.

\subsubsection*{Absorptive sector}

We adopt the in-in formalism in order to accommodate effects which are not time-reversal symmetric~\cite{Galley:2009px,Goldberger:2020fot,Jakobsen:2022psy,Kalin:2022hph}.
In this framework, each dynamical degree of freedom is doubled, with one field evolving forward in time and its duplicate backward in time; in the Keldysh basis, we work with the average and difference of these two fields, respectively denoted with $(\pm)$ superscripts \cite{Keldysh:1964ud}; see \cref{eq:plusminusdef}.

In the classical limit, one only needs the part of the action linear in the minus fields.
This makes it straightforward to re-interpret the in-out part of the action in this language~\cite{Jakobsen:2022psy}:
\begin{align}\label{eq:in-in_action}
  S_{\rm in-in}
  =
  \sum_X
  \int \mathrm{d}x_X
  X^{(-)}(x_X)
  \bigg[\left.
    \frac{\delta S}{\delta X(x_X)}
  \right|_{\{X\}\to \{X^{(+)}\}}
  \bigg]
  \ .
\end{align}
Here, the sum over $X$ covers the three dynamical fields $h_\mn(x)$, $\delta x^\mu(\tau)$ and $\delta\alpha^a(\tau)$ with $x_X=x$ for the graviton and $x_X=\tau$ for the worldline fields.
Clearly, a variation with respect to the minus fields gives the classical equations of motion for the plus fields.
At the boundaries of the evolution, $\tau=\pm\infty$, one imposes that $X^{(-)}$ vanishes. 
On the other hand, $X^{(+)}$ vanishes only initially, while producing physical excitations in the infinite future.
Retarded propagators are employed for propagation from minus to plus fields.

In addition to \cref{eq:in-in_action}, the in-in formalism allows us to write operators which have no in-out analog.
At leading order in the presence of spin (i.e. $\cO(G^5)$), we can include
\begin{equation}\label{eq:AbsorptiveR2}
\begin{aligned}
  \mathcal{L}_{R^2,{\rm abs}}
  =
  E^{(-)}_{AB}& 
  \lambda^{ABCD}_{(E)}
  E_{CD}^{(+)} \\
  &+
  B^{(-)}_{E{AB}}
  \lambda^{ABCD}_{(B)}
  B^{(+)E}{}_{CD}
  \ ,
\end{aligned}
\end{equation}
with $A$, $B$, $C$, etc., denoting indices in the body-fixed frame $\Lambda^\mu_A(\tau)$ of the compact object:
\begin{align}\label{eq:localFrame}
  \Lambda^\mu_A(\tau) \Lambda^\nu_B(\tau) g_\mn(x(\tau)) = \eta_{AB}
  \, .
\end{align}
This represents a new degree of freedom on the worldline, tracking the orientation of the body and therefore intimately connected to rotation~\cite{Liu:2021zxr}.
At the level of individual tensors, $E^{(-)}_{AB}$, for example, is a shorthand for the same object linearized in the minus fields,
\begin{align}
  E^{(-)}_{AB}
  =
  \sum_X
  \int \mathrm{d}x_X
  X^{(-)}(x_X)
  \bigg[\left.
    \frac{\delta E_{AB}}{\delta X(x_X)}
  \right|_{\{X\}\to \{X^{(+)}\}}
  \bigg]
  \ ,
\end{align}
and analogously for $B^{(-)}_{ABC}$.
The same notation as in eq.~\eqref{eq:in-in_action} is used, though now the sum on $X$ should also include $\Lambda^\mu_A(\tau)$.

The tensor structures in \cref{eq:AbsorptiveR2} are
\begin{align}\label{eq:AbsorptiveTensor}
  \lambda_{(X)}^{ABCD}
  =
  S^{BD} 
  &\bigg[
    m^3G^{3}\left(a_{(SX^2;0)}+a_{(SX^2;2)}\chi^2\right)\,
    \eta^{AC}\notag \\
    &\left.+
    \frac{G}{m}\,a_{(S^3X^2;0)}
    (S\cdot S)^{AC}
  \right]
  \ .
\end{align}
In the picture of integrating out the multipoles of the compact body, these are the tensor structures comprising their retarded two-point correlation functions propagating from the plus field to the minus field.
It is now immediate to see why $\mathcal{L}_{R^2,{\rm abs}}$ cannot be written with in-out operators: the structures $\lambda^{ABCD}_{(X)}$ are anti-symmetric in $(AB)\leftrightarrow(CD)$, such that \cref{eq:AbsorptiveR2} would vanish if the $(\pm)$ fields were identical.
Finally, we note in passing that the coefficients $a_{(SX^2;s)}$ and $a_{(S^3 X^2;s)}$ correspond to $H_{\hat s^1}^{E/B}$ and $H_{\hat s^3}^{E/B}$ respectively of ref.~\cite{Saketh:2023bul}.

The two-graviton vertex derived from $\mathcal{L}_{R^2,{\rm abs}}$ reads
\begin{equation}\label{eq:AbsVertex}
  \begin{tikzpicture}[baseline=-5mm,scale=1.2]
  \draw [dotted, thick] (-.75,-0.6) -- (.75,-0.6);
  \draw [photonTest] (-.75,0.) -- (0,-0.6);
  \draw [example1={.55}] (-.75,0.) -- (0,-0.6);
  \draw [photonTest] (0,-0.6) -- (.75,0.);
  \draw [example1={.6}] (0,-0.6) -- (.75,0.);
  \draw [fill=black] (-0.07,-0.67) rectangle (0.07,-.53);
  \node [above left,  font=\footnotesize]  at (-.74,-.1) {$\mu_1\nu_1$};
  \node [above right,  font=\footnotesize] at (.74,-.1)  {$\mu_2\nu_2$};
  \path (0,-0.6) -- (-.75,0) node [pos=.75, below left,  font=\footnotesize, inner sep=2pt] {$k_1$};
  \path (0,-0.6) -- (.75,0)  node [pos=.75, below right, font=\footnotesize, inner sep=2pt] {$k_2$};
  \end{tikzpicture}
  =
  \frac{i\kappa_d^2}{4}
  \mathcal{V}^{\mu_1\nu_1\mu_2\nu_2}(k_1,k_2)
  \ ,
\end{equation}
with
\begin{equation}\label{eq:curlyV}
\begin{aligned}
  &\mathcal{V}^{\mu_1\nu_1\mu_2\nu_2}(k_1,k_2)
  =
  2\pi\delta(k_1\cdot v-k_2\cdot v)
  \\
  &\qquad\qquad\times
  \Pi_2{}^{(\nu_2}{}_{\beta_2}
  \Big[
  \Pi_2{}^{\mu_2)}{}_{\alpha_2}
  \lambda_E^{\alpha_2\beta_2\alpha_1\beta_1}
  \Pi_1{}^{(\mu_1}{}_{\alpha_1}
  \\
  &\qquad\qquad
  +
  K_2{}^{\mu_2)\rho}{}_{\alpha_2}
  \lambda_B^{\alpha_2\beta_2\alpha_1\beta_1}
  K_1{}^{(\mu_1}{}_{\rho\alpha_1}
  \Big]
  \Pi_1{}^{\nu_1)}{}_{\beta_1}
  \ ,
\end{aligned}
\end{equation}
and
\begin{equation}
\begin{aligned}
  \Pi_i{}^\mu{}_\alpha
  &=k_i\cdot v \, \delta^\mu_\alpha - k_{i\alpha} v^\mu
  \ , \\
  K_i{}^{\mu}{}_{\rho\alpha}
  &=k_{i\rho} \delta^\mu_\alpha - k_{i\alpha} \delta^\mu_\rho
  \ .
\end{aligned}
\end{equation}
The aforementioned anti-symmetry of the operators giving rise to this rule translates to a time-reversal anti-symmetry of the rule itself: it picks up a sign upon sending $k_{1}\leftrightarrow -k_{2}$.
In other words, the arrows on the gravitons select the sign of the Feynman rule, and must be respected when using this vertex in Feynman diagrams.
For gravitational-wave scattering, this means they flow from the incoming to the outgoing graviton.

Perturbations of the local frame $\Lambda_{A}^{\mu}(\tau)$ do not matter in our calculation of gravitational-wave scattering at this order.
This is because \cref{eq:AbsVertex} is the only vertex derived from \cref{eq:AbsorptiveR2} which enters the diagrams of this amplitude, and both gravitons are already supplied by the curvature tensors in the action.
In \Cref{sec:Absorption}, however, we also show how $\mathcal{L}_{R^2,{\rm abs}}$ gives rise to absorption of mass and spin in a two-body encounter.
In this context, one also needs vertices with two gravitons and either one worldline or one oscillator fluctuation; see \Cref{app:absComputations}.

As we will see, the values of the coefficients appearing in \cref{eq:AbsorptiveTensor} will be fixed by matching to the absorptive portion of the BHPT result.
It is this matching which permits us to interpret $\mathcal{L}_{R^2,{\rm abs}}$ as the absorptive part of the action.

\section{Matching WQFT to BHPT}\label{sec:Matching}

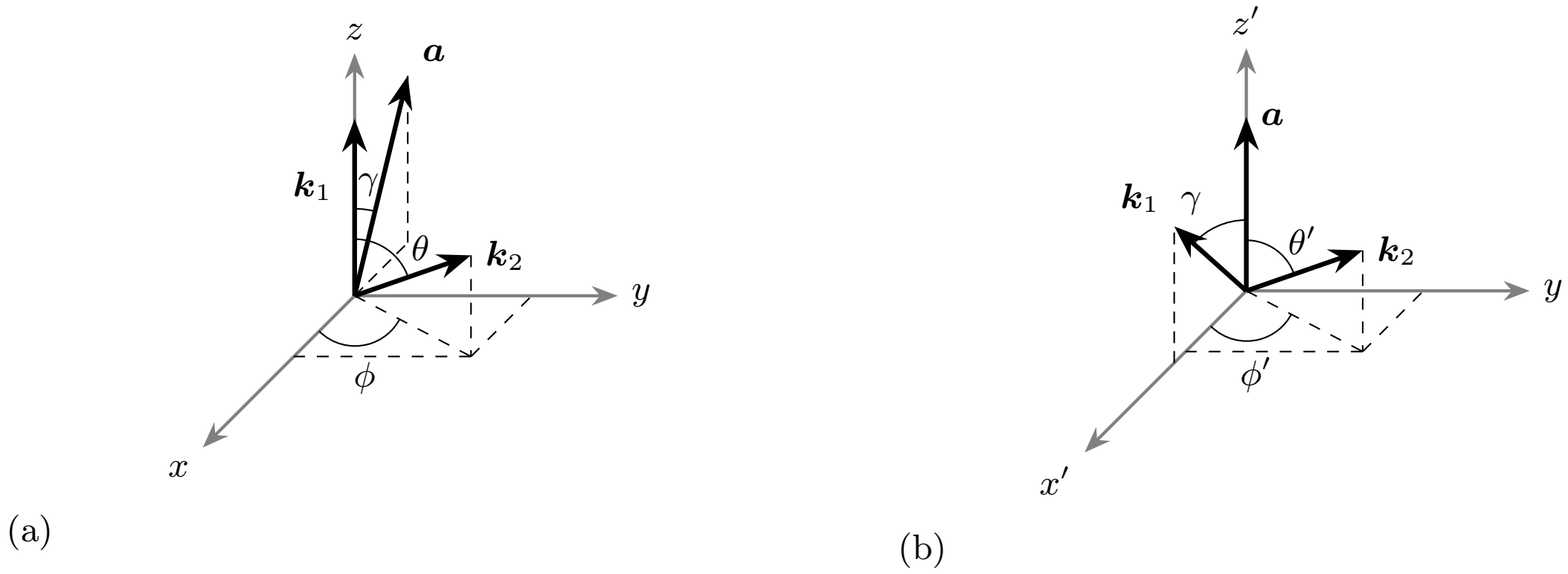
\begin{figure*}[htbp]
  \begin{minipage}{0.4\textwidth}
  \centering
  \begin{tikzpicture}[scale=0.9, >=Stealth]
    \coordinate (O) at (0,0);
    \coordinate (ex) at (-0.75,-0.75);
    \coordinate (ey) at (1,0);
    \coordinate (ez) at (0,1);
    \draw[->,thick,color=black!50] (O) -- ($(O)+2.40*(ez)$) node[above] {${\color{black}z}$};
    \draw[->,thick,color=black!50] (O) -- ($(O)+2.60*(ey)$) node[right] {${\color{black}y}$};
    \draw[->,thick,color=black!50] (O) -- ($(O)+2.00*(ex)$) node[below left] {${\color{black}x}$};
    \coordinate (Qone) at ($(O)+1.75*(ez)$);
    \coordinate (Qtwo) at ($(O)+0.80*(ex)+1.75*(ey)+1.00*(ez)$);
    \coordinate (Ptwo) at ($(O)+0.80*(ex)+1.75*(ey)$);
    \coordinate (Px) at ($(O)+0.80*(ex)$);
    \coordinate (Py) at ($(O)+1.75*(ey)$);
    \coordinate (Avec) at ($(O)-0.70*(ex)+1.65*(ez)$);
    \coordinate (Aproj) at ($(O)-0.70*(ex)$);
    \draw[dashed] (Qtwo) -- (Ptwo);
    \draw[dashed] (O) -- (Ptwo);
    \draw[dashed] (Ptwo) -- (Px);
    \draw[dashed] (Ptwo) -- (Py);
    \draw[densely dashed,thin] (O) -- (Aproj);
    \draw[densely dashed,thin] (Aproj) -- (Avec);
    \draw[->,very thick] (O) -- (Qone)
      node[pos=0.62,left,xshift=-2pt] {$\boldsymbol{k}_1$};
    \draw[->,very thick] (O) -- (Qtwo) node[right] {$\boldsymbol{k}_2$};
    \draw[->,very thick] (O) -- (Avec) node[above right] {$\boldsymbol{a}$};
    \draw (0,0.56) arc[start angle=90,end angle=19.2,radius=0.56]
      node[pos=0.45,right,xshift=2pt] {$\,\theta$};
    \draw (-0.35,-0.35) arc[start angle=-135,end angle=-27.6,radius=0.50]
      node[pos=0.52,below,yshift=-1pt] {$\phi$};
    \draw (0,0.86) arc[start angle=90,end angle=76.4,radius=0.86]
      node[pos=0.3,above,xshift=2pt] {$\gamma$};
  \end{tikzpicture}
  \subcaption{\label{fig:conventions_WQFT}}
\end{minipage}
\qquad
\begin{minipage}{0.4\textwidth}
  \centering
  \begin{tikzpicture}[scale=0.9, >=Stealth]
    \coordinate (O) at (0,0);
    \coordinate (ex) at (-0.75,-0.75);
    \coordinate (ey) at (1,0);
    \coordinate (ez) at (0,1);
    \draw[->,thick,color=black!50] (O) -- ($(O)+2.40*(ez)$) node[above] {${\color{black}z^\prime}$};
    \draw[->,thick,color=black!50] (O) -- ($(O)+2.80*(ey)$) node[right] {${\color{black}y}$};
    \draw[->,thick,color=black!50] (O) -- ($(O)+2.13*(ex)$) node[below left] {${\color{black}x^\prime}$};
    \coordinate (Qone) at ($(O)+0.95*(ex)+1.35*(ez)$);
    \coordinate (Qtwo) at ($(O)+0.80*(ex)+1.75*(ey)+1.00*(ez)$);
    \coordinate (Ptwo) at ($(O)+0.80*(ex)+1.75*(ey)$);
    \coordinate (Px) at ($(O)+0.80*(ex)$);
    \coordinate (Py) at ($(O)+1.75*(ey)$);
    \coordinate (Avec) at ($(O)+1.72*(ez)$);
    \coordinate (QoneProj) at ($(O)+0.95*(ex)$);
    \draw[dashed] (Qtwo) -- (Ptwo);
    \draw[dashed] (O) -- (Ptwo);
    \draw[dashed] (Ptwo) -- (Px);
    \draw[dashed] (Ptwo) -- (Py);
    \draw[densely dashed,thin] (QoneProj) -- (Qone);
    \draw[->,very thick] (O) -- (Qone) node[above left] {$\boldsymbol{k}_1$};
    \draw[->,very thick] (O) -- (Qtwo) node[right] {$\boldsymbol{k}_2$};
    \draw[->,very thick] (O) -- (Avec) node[right] {$\boldsymbol{a}$};
    \draw (0,0.50) arc[start angle=90,end angle=19.2,radius=0.50]
      node[pos=0.05,right,xshift=6.5pt] {$\theta'$};
    \draw (-0.35,-0.35) arc[start angle=-135,end angle=-27.6,radius=0.50]
      node[pos=0.52,below,yshift=-1pt] {$\phi'$};
    \draw (0,0.70) arc[start angle=90,end angle=138.2,radius=0.70]
      node[pos=0.55,above left] {$\gamma$};
  \end{tikzpicture}
  \subcaption{\label{fig:bhpt-frame}}
\end{minipage}
\caption{The coordinate systems used to express (a) the WQFT and (b) the BHPT results, referred to respectively as the ``WQFT frame" and the ``BHPT frame".}
\end{figure*}

The above analysis of the required operators in our action translates to concrete precision goals for the amplitude for gravitational-wave scattering: namely, the action we have established is sufficient for completely evaluating the amplitude to $\cO(G^5)$.
We seek to match the WQFT computation performed to this order to the analogous result from BHPT.

\subsection{WQFT framework}

Concretely, the amplitude is the $S$-matrix element for single-graviton incoming and outgoing states in the presence of a spinning source:
\begin{equation}\label{eq:Amplitude}
\begin{aligned}
  \langle &k_{2},h_{2}|i\hat{T}|k_{1},h_{1}\rangle \\
  &=2\pi\,\delta(v\cdot k_{2}-v\cdot k_{1})i\cM_{h_{1},h_{2}}(\ePM,\chi,\omega;\theta,\phi,\gamma),
\end{aligned}
\end{equation}
where $\hat{S}=1+i\hat{T}$.
Recall that we work with the in-in formalism, such that causality flows from the incoming to the outgoing graviton.
To explain the parameters on which the amplitude depends, we import the kinematics of ref.~\cite{Bautista:2026qse} for the incoming and outgoing graviton momenta and polarizations:
\begin{equation}\label{eq:GravitonKinematics}
\begin{aligned}
  k_{1}^{\mu}
  &= \omega(1,0,0,1),
  \qquad
  \epsilon_{1}^{\sigma_{1},\mu}
  = \frac{1}{\sqrt{2}}(0,1,i\sigma_{1},0),
  \\
  k_{2}^{\mu}
  &= \omega(1,s_{\theta}c_{\phi},s_{\theta}s_{\phi},c_{\theta}),
  \\
  \epsilon_{2}^{\sigma_{2},\mu}
  &= \frac{1}{\sqrt{2}}\left(
      0,
      c_{\theta}c_{\phi}-i\sigma_{2}s_{\phi},
      c_{\theta}s_{\phi}+i\sigma_{2}c_{\phi},
      -s_{\theta}
      \right),
\end{aligned}
\end{equation}
abbreviating $s_{\beta}\equiv\sin\beta$ and $c_{\beta}\equiv\cos\beta$ for any angle $\beta$, and the signs of the helicities are carried by $\sigma_{i}\equiv h_{i}/2$.
Some quantities depend only on the relative sign of the helicities, in which case they are a function of $\sigma\equiv\sigma_{1}\sigma_{2}$.
Graviton 2 is outgoing, so it is the conjugate of its polarization printed above which enters the amplitude.
The scattering angle $\theta$ may range from $0$ to $\pi$, while the azimuthal angle $\phi\in[0,2\pi)$.

From the graviton energy $\omega$ defined in \cref{eq:GravitonKinematics}, we construct the dimensionless quantity
\begin{align}
  \ePM\equiv2Gm\omega,
\end{align}
which emerges as the perturbative PM parameter; due to spin effects, this is not equivalent to a loop counting parameter.
The amplitude's soft pole
\begin{equation}\label{eq:AmplitudeSoftPole}
\begin{aligned}
  i\cM_{h_{1},h_{2}}(\ePM,\chi&,\omega;\theta,\phi,\gamma) \\
  &=\frac{i\cM_{h_{1},h_{2}}(\ePM,\chi;\theta,\phi,\gamma)}{\omega}
\end{aligned}
\end{equation}
is its only dependence on $\omega$ separate from $\ePM$.

Moving on, the remaining variables $\chi$ and $\gamma$ are related to the body's spin.
We have defined the former in \cref{eq:ChiParameter}.
Working in the rest frame of the compact body, $v^\mu=(1,\boldsymbol{0})$, we choose coordinates that place the (mass-rescaled) spin vector in the $x-z$ plane,
\begin{align}\label{eq:SpinCoords}
  a^{\mu}&= |a|(0,-s_{\gamma},0,c_{\gamma}),
\end{align}
for $\gamma\in[0,\pi)$.
This vector captures the spatial components of the asymptotic spin tensor,
\begin{align}\label{eq:RingRadiusDef}
  S^{\mu\nu}_{0}=m\epsilon^{\mu\nu\alpha\beta}v_{\alpha}a_{\beta}.
\end{align}
Notice that substituting \cref{eq:RingRadiusDef} into \cref{eq:ChiParameter} leads to
\begin{align}
  \chi=\frac{|a|}{Gm},
\end{align}
such that the magnitude of the spin vector is not a new parameter.

\Cref{fig:conventions_WQFT} depicts the coordinate system referred to by \cref{eq:GravitonKinematics,eq:SpinCoords}, which we dub the WQFT frame.

In addition to each loop order in a Feynman diagrammatic expansion, and as argued above, each power of spin also increases the PM relevance of a contribution to the amplitude by one order.
Capturing all effects up to $\cO(G^5)$ thus entails computing all corrections where the loop order $l$ and the spin order $s$ obey $l+s\leq4$, as  summarized in \cref{tab:AmplitudeTargets}.

\begin{table*}[htbp]
  \begin{minipage}{0.4\textwidth}
  \begin{tabular}{|c||c|c|c|c|c|c}
    \hline
    \backslashbox{loops}{spin} & 0 & 1 & 2 & 3 & 4 \\
    \hline
    \hline
    \rule{0pt}{3.5ex}0 & $\color{teal}\,G^{1}\,$ & $\color{teal}\,G^{2}\,$ & $\color{teal}\,G^{3}\,$ & $\color{teal}\,G^{4}\,$ & $\color{teal}\,G^{5}\,$ \\[1ex]
    \hline
    \rule{0pt}{3.5ex}1 & $\color{teal}\,G^{2}\,$ & $\color{Plum}\,G^{3}\,$ & $\color{Plum}\,G^{4}\,$ & $\color{Plum}\,G^{5}\,$ & \\[1ex]
    \hline
    \rule{0pt}{3.5ex}2 & $\color{teal}\,G^{3}\,$ & $\color{Red}\,G^{4}\,$ & $\color{Red}\,G^{5}\,$ & & \\[1ex]
    \hline
    \rule{0pt}{3.5ex}3 & $\color{teal}\,G^{4}\,$ & $\color{Red}\,G^{5}\,$ & & & \\[1ex]
    \hline
    \rule{0pt}{3.5ex}4 & $\color{teal}\,G^{5}\,$ & & & & \\[1ex]
    \hline
  \end{tabular} \\[2ex]
  $\cO(R)$
  \end{minipage}
  \begin{minipage}{0.4\textwidth}
  \begin{tabular}{|c||c|c|c|c|c|c}
    \hline
    \backslashbox{loops}{spin} & 0 & 1 & 2 & 3 & 4 \\
    \hline
    \hline
    \rule{0pt}{3.5ex}0 & $\color{teal}\,G^{5}\,$ & \cellcolor{ForestGreen!20} $\color{teal}\,G^{5}\,$ & $\color{Red}\,G^{5}\,$ & \cellcolor{ForestGreen!20} $\color{teal}\,G^{5}\,$ & $\color{teal}\,G^{5}\,$ \\[1ex]
    \hline
  \end{tabular} \\[23ex]
  $\cO(R^{2})$
  \end{minipage}
  \caption{\label{tab:AmplitudeTargets}The contributions to the gravitational-wave scattering amplitude needed to match to BHPT up to $\cO(G^5)$ coming from \textit{left:} linear- and \textit{right:} quadratic-in-curvature operators. Blue entries have been computed and matched to BHPT \cite{Bautista:2022wjf,Saketh:2023bul,Ivanov:2026icp,Bautista:2026qse,Bjerrum-Bohr:2026fhs,Bautista:2026fcp,Brunello:2026rdk,Brunello:2026lzf,Correia:2026utp}; those in purple are known but not matched \cite{Akpinar:2025byi}; we compute red ones for the first time here. Light-green shading indicates absorptive contributions, while white cells are conservative effects.}
\end{table*}

Beyond reproducing known results at tree level \cite{Arkani-Hamed:2017jhn,Guevara:2018wpp,Aoude:2020onz,Aoude:2022trd,Bautista:2022wjf} (including absorption \cite{Porto:2007qi,Goldberger:2020fot,Saketh:2023bul,Bautista:2024emt}), one-loop \cite{Akpinar:2025byi}, and spinless orders \cite{Bjerrum-Bohr:2025bqg,Bautista:2026qse,Ivanov:2025ozg,Bjerrum-Bohr:2026fhs,Bautista:2026fcp,Brunello:2026rdk,Brunello:2026lzf,Correia:2026utp}, we evaluate all missing pieces for the completion of the amplitude, as well as the matrix element of the Magnus operator (i.e. the Magnusian),
\begin{equation}\label{eq:MagnusAmplitude}
\begin{aligned}
  \langle &k_{2},h_{2}|i\hat{N}|k_{1},h_{1}\rangle \\
  &=2\pi\,\delta(v\cdot k_{2}-v\cdot k_{1})\frac{1}{\omega}iN_{h_{1},h_{2}}(\ePM,\chi;\theta,\phi,\gamma),
\end{aligned}
\end{equation}
up to $\cO(G^5)$.
This latter quantity enables an efficient matching to BHPT, and is defined through the exponential representation of the $S$-matrix \cite{Damgaard:2021ipf,Kim:2024svw,Kim:2025gis,Brandhuber:2025igz,Gonzo:2026yha,Brandhuber:2026dkd},
\begin{align}
  \hat{S}=e^{i\hat{N}}.
\end{align}
Note that we have explicitly factored out the soft pole from the Magnusian in \cref{eq:MagnusAmplitude} for convenience.



In order to compare \cref{eq:MagnusAmplitude} to the result obtained in the BHPT framework, we follow the strategy presented in refs.~\cite{Bautista:2021wfy,Bautista:2022wjf}, involving projecting the Magnusian onto a basis of $s=-2$ spin-weighted spherical harmonics:
%
\begin{align}
  &N_{h_1,h_2}(\ePM,\chi;\theta,\phi,\gamma)\label{eq:Delta-expansion} \\
  &=(-1)^{m-2}
  \sum_{\ell=2}^{\infty}\sum_{m=-\ell}^{\ell}
  N_{h_1,h_2;\ell m}(\ePM,\chi;\gamma)\,
  \Ysw{-2}{\ell}{m}(\theta_{\sigma},\phi)\,,\notag
\end{align}
where
\begin{align*}
  \theta_{\sigma}
  &=
  \begin{cases}
    \theta, & \sigma=1, \\
    \pi-\theta, & \sigma=-1.
  \end{cases}
\end{align*}
Inverting \cref{eq:Delta-expansion} extracts the modes of the projection,
\begin{align}
  &N_{h_1,h_2;\ell m}(\ePM,\chi,\gamma)\label{eq:N-projection} \\
  &=(-1)^{m-2}
  \int \mathrm{d} \Omega\,
  {}_{-2}\overline{Y}_{\ell m}(\theta_{\sigma},\phi) 
  N_{h_1,h_2}(\ePM,\chi;\theta,\phi,\gamma)\,,\notag
\end{align}
which makes use of the orthonormality of the spin-weighted spherical harmonics.

The WQFT frame defined above is particularly convenient for evaluating \cref{eq:N-projection}, since the magnitude of the momentum transfer is independent of the azimuthal angle $\phi$:
\begin{equation}\label{eq:TransferMomentum}
  q^2
  =
  (k_{2}-k_{1})^{2}
  = -4\omega^{2}\sin^2(\theta/2)
  \equiv -4\omega^{2}x^{2}.
\end{equation}
This simplifies the azimuthal integration in the partial-wave projection, particularly when $q^2$ appears in denominators or the arguments of special functions.
Dependence on $\phi$ can still arise through Lorentz contractions with the spin, e.g.
\begin{equation}\label{eq:adotk2}
  a\cdot k_{2}
  = \frac{\chi\ePM}{2}
    \left(
      s_\gamma\,s_\theta\,c_\phi
      - c_\gamma\,c_\theta
    \right)\,.
\end{equation}
However, since the WQFT computation is arranged polynomially in $a^\mu$, this dependence appears polynomially rather than in denominators or special functions.
Integrating over $\phi$ is therefore elementary.




\subsection{BHPT framework}
The BHPT amplitudes are conventionally evaluated in the $(x',y,z')$ (BHPT) frame shown in \cref{fig:bhpt-frame}. In this frame, the black hole is at rest, with its spin aligned with the $z'$ axis. The incoming wave propagates in the $x'-z'$ plane at an angle $\gamma$ to the $z'$ axis, and its helicity is fixed to $h_{1}=+2$. The outgoing wave propagates in a generic direction, parametrized by the angles $(\theta',\phi')$, and may have either positive or negative helicity.

BHPT encodes information about the scattering in the phase shift $\delta$ and absorption factor $\eta$, in the combination
\begin{equation}\label{eq:BHPTScatteringInfo}
\begin{aligned}
 & {}_{-2}\mathcal{F}^{P}_{\ell^{\prime}m^{\prime}}(\ePM,\chi) \\
  &\equiv{}_{-2}\delta^{P}_{\ell^{\prime}m^{\prime}}(\ePM,\chi)-\frac{i}{2}\log[{}_{-2}\eta_{\ell^{\prime}m^{\prime}}(\ePM,\chi)]\,,
\end{aligned}
\end{equation}
whose explicit form we obtain up to order $\ePM^5$ following ref.~\cite{Bautista:2023sdf}; see Appendix \ref{app:BHPTResults}. 
Here $P$ is a parity label and we have primed the angular momentum numbers to indicate modes in the BHPT frame.
In this frame, the projection is on spin-weighted spheroidal harmonics $\Ssw{-2}{\ell'}{m'}(\alpha',\beta',|a|\omega)$,\footnote{We write the last argument of these functions as $|a|\omega$ rather than the equivalent $\ePM\chi/2$ to reduce visual clutter.} such that helicity preserving ($h_{1}=h_{2}=2$) and reversing ($h_{1}=-h_{2}=2$) amplitudes are \cite{Dolan:2008kf}
\begin{align}
  f&(\ePM,\chi;\theta^\prime,\phi^\prime,\gamma)\label{eq:fBHPTAmp} \\
  &=\frac{2\pi}{i\omega}\sum_{\ell',m'}
  \sum_{P=\pm 1} \,
  \left(e^{2i{}_{-2}\mathcal{F}^P_{\ell' m'}}-1\right)\notag \\
  &\times
  \Ssw{-2}{\ell'}{m'}(\gamma,0,|a|\omega)\,
  \Ssw{-2}{\ell'}{m'}(\theta',\phi',|a|\omega)\notag \\[2ex]
  g&(\ePM,\chi;\theta^\prime,\phi^\prime,\gamma)\label{eq:gBHPTAmp} \\
  &=\frac{2\pi}{i\omega}\sum_{\ell',m'}
  \sum_{P=\pm 1} \,
  P(-1)^{\ell^\prime}\left(e^{2i{}_{-2}\mathcal{F}^P_{\ell' m'}}-1\right)\notag \\
  &\times
  \Ssw{-2}{\ell'}{m'}(\gamma,0,|a|\omega)\,
  \Ssw{-2}{\ell'}{m'}(\pi-\theta',\phi',|a|\omega)\notag
\end{align}
respectively, with the same bounds on the sums as in \cref{eq:Delta-expansion}.
Spin-weighted spheroidal harmonics are the solutions to the angular part of the Teukolsky equation.
Their closed forms are not known, but they reduce to spin-weighted spherical harmonics when $|a|\omega\rightarrow0$, and otherwise can be expanded in a basis of the latter \cite{Dolan:2008kf}.

While $f$ and $g$ should be matched to \cref{eq:Amplitude}, we have proven in ref.~\cite{Bautista:2026qse} that \cref{eq:N-projection} can be readily matched to \cref{eq:BHPTScatteringInfo} in the absence of spin.
The situation is not so immediate in the presence of spin, in part due to the more complicated angular and helicity dependence of the Magnusian, but also because of the two different frames in \cref{fig:bhpt-frame}.
Nevertheless, we observe that a comparison can be made between the Magnusian and \cref{eq:BHPTScatteringInfo} when spin is involved too.

Elaborating on this, define the BHPT Magnusian:
\begin{equation}\label{eq:bhpt-expansion}
\begin{aligned}
  \Delta_{h_1,h_2}&(\ePM,\chi;\theta',\phi',\gamma) \\
  &:=
  4\pi
  \sum_{\ell',m'}
  \sum_{P=\pm 1} \zeta_{\sigma;\ell'}^{P}\,
  {}_{-2}\mathcal{F}^P_{\ell' m'}(\ePM,\chi) \\
  &\times
  \Ssw{-2}{\ell'}{m'}(\gamma,0,|a|\omega)\,
  \Ssw{-2}{\ell'}{m'}(\theta'_{\sigma},\phi',|a|\omega)\,,
\end{aligned}
\end{equation}
with
\begin{align}
  (\zeta_{\sigma;\ell^{\prime}}^{P},\theta^{\prime}_{\sigma})
  &=
  \begin{cases}
    (1,\theta^{\prime}), & \sigma=1, \\
    (P(-1)^{\ell^{\prime}},\pi-\theta^{\prime}), & \sigma=-1.
  \end{cases}
\end{align}
In contrast to \cref{eq:fBHPTAmp,eq:gBHPTAmp}, the presence of $\zeta_{\sigma;\ell^{\prime}}^{P}$ combined with the independence of the absorption probability of the parity label $P$ means that the helicity reversing BHPT Magnusian is conservative to all orders in perturbation theory.

\subsection{Frame rotation and matching}

Modes of the BHPT Magnusian which are suitable for comparison with those in \cref{eq:N-projection} are obtained in two steps.
First, we project \cref{eq:bhpt-expansion} onto a basis of spin-weighted spherical harmonics, producing the modes
\begin{equation}
  \begin{split}
  &\Delta_{h_1,h_2;\ell' m'}(\ePM,\chi;\gamma)
  \\&=
  \int \text{d}\Omega'\,
  {}_{-2}\overline{Y}_{\ell' m'}(\theta'_{\sigma},\phi')\,
  \Delta_{h_1,h_2}(\ePM,\chi;\theta',\phi',\gamma).
  \label{eq:Delta-lm}
  \end{split}
\end{equation}
In terms of these modes, we obtain a second representation of the BHPT Magnusian:
\begin{align}
  \Delta_{h_1,h_2}&(\ePM,\chi;\theta',\phi',\gamma) \\
  &=\sum_{\ell,m}\Delta_{h_1,h_2;\ell' m'}(\ePM,\chi;\gamma)Y_{\ell' m'}(\theta'_{\sigma},\phi').\notag
\end{align}
Next, we must rotate this from the primed BHPT frame, by an angle $\gamma$ around the $y$-axis, into the unprimed WQFT frame, so that
\[
  (\theta',\phi')\longrightarrow(\theta,\phi).
\]
This would land us on
\begin{align}
  \Delta_{h_1,h_2}&(\ePM,\chi;\theta,\phi,\gamma)\label{eq:Delta-prime-expansion} \\
  &=
  \sum_{\ell,m}
  \Delta_{h_1,h_2;\ell m}(\ePM,\chi;\gamma)\,
  \Ysw{-2}{\ell}{m}(\theta_\sigma,\phi),\notag
\end{align}
with modes in the same frame as those of \cref{eq:N-projection}.
We are left with the question of how we can reach these modes given those defined by \cref{eq:Delta-lm}, to which we have access from the BHPT Magnusian, \cref{eq:bhpt-expansion}.

To answer this question, we follow Appendix B of ref.~\cite{Boyle:2013nka}.
Whereas the BHPT Magnusian and the spin-weighted spherical harmonics transform with spin weight 2 under rotations, the modes of the projection transform as scalars.
The desired rotation by an angle $\gamma$ about the $y$-axis is produced by the Wigner-$D$ matrices,
\begin{equation}
\begin{aligned}
  \Dwig{\ell\,*}{m m'} \,(\gamma)&\equiv\Dwig{\ell\,*}{m m'} \,(0,\gamma,0) \\
  &=(-1)^{m'}\sqrt{\frac{4\pi}{2\ell+1}}{}_{-m'}Y_{\ell m}(\gamma,0)\,,
\end{aligned}
\end{equation}
such that
\begin{equation}\label{eq:boxed-Delta-prime}
\begin{aligned}
  &\Delta_{h_1,h_2;\ell m}(\ePM,\chi;\gamma)=
  \sum_{m'}
  \Dwig{\ell\,*}{m' m}\,(\gamma) \\
  &\times
  \int \text{d}\Omega'\,
  {}_{-2}\overline{Y}_{\ell m'}(\theta'_{\sigma},\phi')\,
  \Delta_{h_1,h_2}(\ePM,\chi;\theta',\phi',\gamma)\,,
\end{aligned}
\end{equation}
where the integral in the second line defines the modes in the BHPT frame, \cref{eq:Delta-lm}.

We can now extract from \cref{eq:bhpt-expansion} modes which are comparable to  \cref{eq:N-projection}: matching WQFT to BHPT means
\begin{equation}
  \boxed{
    N_{h_1,h_2;\ell m}(\ePM,\chi;\gamma)
    =
    2\pi\,\Delta_{h_1,h_2;\ell m}(\ePM,\chi;\gamma)
  }\,\,.
  \label{eq:Nprime-Delta-prime}
\end{equation}
Both the factor of $2\pi$ here as well as the $m$-alternating   sign in \cref{eq:Delta-expansion} are determined empirically already from matching at 1PM.
Then, where free Wilson coefficients exist in the Magnusian, imposing this equation uniquely fixes their values.

Finally, let us close this section by noting that all quantities mentioned are expanded in $\ePM$:
\begin{align}\label{eq:PMExpansion}
  X(\ePM,\chi;\Omega)=\sum_{n=1}^{\infty}\ePM^{n}X^{(n)}(\chi;\Omega)\,,
\end{align}
where $X$ stands for any amplitude or mode thereof, and $\Omega$ stands in for its angular dependence.
On the other hand, the parameter $\chi$ is not taken to be perturbative, but rather lies in the physical region $\chi\le1$.


\section{Computing the Magnusian with spin}\label{sec:Computation}

Up until now, we have discussed how the WQFT and BHPT frameworks are modified to account for spin effects.
Here we explore WQFT computational aspects, both relating to spin effects as well as to the spinless Magnusian at four-loop order.

\subsection{Diagrammatics}

Introducing spin effects brings about new vertices involving bosonic oscillator excitations and direct couplings between the worldline and two gravitons, proliferating the number of diagrams.
Nevertheless, the diagram generation method of ref.~\cite{Bautista:2026qse} can be employed with only one adjustment:
replacing diagrams with a worldline propagator with an effective two-graviton vertex accounting also for new spin-induced vertices:
\begin{equation}\label{eq:TwoGravEffectiveVertex}
\begin{aligned}
  \begin{tikzpicture}[baseline=-5mm,scale=1.2]
    \draw [dotted, thick] (-.75, -0.6) -- (.75, -0.6);
    \draw [photonTest] (0,-0.6) -- (.75,0.);
    \draw [photonTest] (0,-0.6) -- (-.75,0.);
    \draw [fill=white, thick] (-0.18,-0.78) rectangle (.18,-.42);
    \node[cross=4.2pt] at (0,-0.6) {};
  \end{tikzpicture}
  \,\,
  &=
  \,\,
  \begin{tikzpicture}[baseline=-5mm,scale=1.2]
    \draw [dotted, thick] (-.75, -0.6) -- (.75, -0.6);
    \draw [photonTest] (0,-0.6) -- (.75,0.);
    \draw [photonTest] (0,-0.6) -- (-.75,0.);
    \draw [fill=black] (0.,-0.6) circle (.067);
  \end{tikzpicture}
  \,\,
  +
  \,\,
  \begin{tikzpicture}[baseline=-5mm,scale=1.2]
    \draw [dotted, thick] (-.75, -0.6) -- (-.5, -0.6);
    \draw [dotted, thick] (.5, -0.6) -- (.75, -0.6);
    \draw [zParticle, thick] (-.5, -0.6) node [below] {$\delta x(-\omega)$} -- (.5, -0.6);
    \draw [photonTest] (.5,-0.6) node [below] {$\delta x(\omega)$} -- (.75,0.);
    \draw [photonTest] (-.5,-0.6) -- (-.75,0.);
    \draw [fill=black] (-0.5,-0.6) circle (.067);
    \draw [fill=black] (0.5,-0.6) circle (.067);
  \end{tikzpicture} \\
  &\!\!\!\!\!\!\!\!+
  \begin{tikzpicture}[baseline=-5mm,scale=1.2]
    \draw [dotted, thick] (-.75, -0.6) -- (-.5, -0.6);
    \draw [dotted, thick] (.5, -0.6) -- (.75, -0.6);
    \draw [zParticle, thick] (-.5, -0.6) node [below] {$\delta\bar{\alpha}(-\omega)$} -- (.5, -0.6);
    \draw [photonTest] (.5,-0.6) node [below] {$\delta\alpha(\omega)$} -- (.75,0.);
    \draw [photonTest] (-.5,-0.6) -- (-.75,0.);
    \draw [fill=black] (-0.5,-0.6) circle (.067);
    \draw [fill=black] (0.5,-0.6) circle (.067);
  \end{tikzpicture}
  \,\,
  +
  \,\,
  \begin{tikzpicture}[baseline=-5mm,scale=1.2]
    \draw [dotted, thick] (-.75, -0.6) -- (-.5, -0.6);
    \draw [dotted, thick] (.5, -0.6) -- (.75, -0.6);
    \draw [zParticle, thick] (-.5, -0.6) node [below] {$\delta\alpha(-\omega)$} -- (.5, -0.6);
    \draw [photonTest] (.5,-0.6) node [below] {$\delta\bar{\alpha}(\omega)$} -- (.75,0.);
    \draw [photonTest] (-.5,-0.6) -- (-.75,0.);
    \draw [fill=black] (-0.5,-0.6) circle (.067);
    \draw [fill=black] (0.5,-0.6) circle (.067);
  \end{tikzpicture}
  \,\,
  .
\end{aligned}
\end{equation}
Energy always flows from left to right in these diagrams.
Combining the last two diagrams here allows one to eliminate bare oscillators for a spin tensor through \cref{eq:SpinTensor}.
The first diagram on the right-hand side includes contributions from the operators in \cref{sec:R2Ops}; one must be careful to respect the causal flow of the absorptive vertex \cref{eq:AbsVertex}.

Incorporating this, the geometric series generating spinless diagrams in ref.~\cite{Bautista:2026qse} is augmented to
\begin{equation}\label{eq:res_com}
  \begin{tikzpicture}[baseline=-4mm,scale=1.2]
    \draw [dotted, thick] (-.75, -0.6) -- (.75, -0.6);
    \draw [photonTest] (0,-0.6) -- (.75,0.);
    \draw [photonTest] (0,-0.6) -- (-.75,0.);
    \draw [fill=black!60, thick] (0,-0.6) circle (0.2cm);;
  \end{tikzpicture}
  \ \ =\ \ 
  \frac{
    \begin{tikzpicture}[baseline=1mm,xscale=-1]
    \draw [dotted, thick] (0,0) -- (1.2,0);
    \draw [photonTest] (1.2,.3) -- (0,.3);
  \end{tikzpicture}
    \rule[-0.4cm]{0pt}{0cm}
  }{
  \rule{0pt}{.7cm}
  1\,\, -\,\,
  \begin{tikzpicture}[baseline=-4mm,xscale=-1]
    \draw [dotted, thick] (-.75, -0.6) -- (.75, -0.6);
    \draw [photonTest] (0,-0.6) -- (.75,0.);
    \draw [photonTest] (0,-0.6) -- (-.75,0.);
    \draw [fill=white, thick] (0,-0.6) circle (0.2cm);
    \node[cross=3.5pt] at (0,-0.6) {};
  \end{tikzpicture}
  \,\,-\,\,
  \begin{tikzpicture}[baseline=-4mm,xscale=-1]
    \draw [dotted, thick] (-.75, -0.6) -- (.75, -0.6);
    \draw [photonTest] (0,-0.6) -- (.75,0.);
    \draw [photonTest] (0,-0.6) -- (-.75,0.);
    \draw [fill=white, thick] (-0.18,-0.78) rectangle (.18,-.42);
    \node[cross=3.5pt] at (0,-0.6) {};
  \end{tikzpicture}
  }
  \ \ .
\end{equation}
In contrast to the new effective vertex, that denoted by a circle and a cross represents all ways of coupling two gravitons to the worldline without exciting any worldline/oscillator propagators; this object has already appeared in ref.~\cite{Bautista:2026qse}, and reads
\begin{align}\label{eq:MetricEffectiveVertex}
  \begin{tikzpicture}[baseline=-4mm,scale=1.2]
    \draw [dotted, thick] (-.75, -0.6) -- (.75, -0.6);
    \draw [photonTest] (0,-0.6) -- (.75,0.);
    \draw [photonTest] (0,-0.6) -- (-.75,0.);
    \draw [fill=white, thick] (0,-0.6) circle (0.2cm);
    \node[cross=3.5pt] at (0,-0.6) {};
  \end{tikzpicture}
  \ \ 
  =
  \ \ 
  \sum_{n=1}^{\infty}\frac{1}{n!}
  \,
  \begin{tikzpicture}[baseline=-1mm,xscale=-1]
    \draw [dotted, thick] (-.75, -0.6) -- (.75, -0.6);
    \draw [photonTest] (0,0.4) -- (-.75,0.4);
    \draw [photonTest] (0,0.4) -- (.75,0.4);
    \draw [fill] (0,0.4) circle (.08);
    \draw [photonTest] (0,0.4) -- (-0.4,-0.4);
    \draw [fill=black!60, thick] (-0.4,-0.6) circle (0.2cm);
    \draw [photonTest] (0,0.4) -- (0.4,-0.4);
    \draw [fill=black!60, thick] (0.4,-0.6) circle (0.2cm);
    \node at (-.04,-0.3) {$\cdots$};
    \node at (0.02,-1.05) {$\underbrace{\hphantom{hej hej he}}_n$};
  \end{tikzpicture}
  \,\,.
\end{align}
Whereas \cref{eq:TwoGravEffectiveVertex} is tree-level exact, the circular effective vertex involves the all-loop one-graviton vertex, encoding the all-order metric of the compact body represented by the worldline.

With \cref{eq:TwoGravEffectiveVertex,eq:MetricEffectiveVertex} at hand, all spinless diagrams ---e.g. those in ref.~\cite{Bautista:2026qse}--- suffice for the spinning case too upon swapping worldline propagators for \cref{eq:TwoGravEffectiveVertex}.
As a corollary, any $l$-loop diagram comprises exactly $l+1$ vertices selected from the one-graviton or the new effective two-graviton vertex.
Another feature of this grouping is that \cref{eq:MetricEffectiveVertex} encodes the metric sourced by the compact object, while \cref{eq:TwoGravEffectiveVertex} carries short-range information about the body.
It is also interesting to note the simplicity of diagrams involving \cref{eq:TwoGravEffectiveVertex}, which effectively factorize into incoming and outgoing dynamics.
In such diagrams, the two external momenta can never appear in the same loop integral, meaning all such integrals are simply constants and the diagram can only depend polynomially on the scattering angle.

With or without spin, then, the tree-level through four-loop integrands are comprised of 2, 6, 20, 70, and 248 diagrams respectively, so long as the effective vertex in \cref{eq:TwoGravEffectiveVertex} is used in place of its constituents.

It has been observed in ref.~\cite{Brunello:2026rdk} up to $\cO(G^4)$ that diagrams involving a worldline propagator do not contribute in the frame defined by \cref{eq:GravitonKinematics} in the absence of spin.
This observation was proven to hold to all orders in perturbation theory in ref.~\cite{Correia:2026utp}, establishing the irrelevance of recoil operators in the aforementioned frame.
We give in \Cref{app:RecoilProof} a simple, all-order proof based on WQFT diagrammatics that, in this frame, the recoil can enter for the first time at quadratic order in spin.
Additionally, we show that spin recoil, reflected by bosonic-oscillator propagators, enters no sooner than cubic order in spin.
We emphasize that this absence of recoil is purely a consequence of a convenient frame choice, rather than carrying a deeper physical significance; we must include recoil in our computation to construct gauge-invariant and Lorentz-covariant results.

\subsection{Integration}

Spin effects play no role in the master integrals which must be solved, but rather simply increase the maximum rank of our tensor integrals.
While for a spinless compact object the highest rank tensor structure which may appear is four, the calculations here produced up to rank eight in our integrands due to the two indices carried by the spin tensor \cref{eq:SpinTensor}.
These were tensor reduced using a privately modified version of \texttt{OPITeR} \cite{Goode:2024cfy},\footnote{We thank Giacomo Brunello for bringing this package to our attention.} resulting in integrals with a maximum of eight scalar products at two loops, seven scalar products at three loops, and six scalar products at four loops.
Otherwise, master integrals up to three loops are exactly those solved in refs.~\cite{Bautista:2026qse,Bautista:2026fcp}, whose values we import here.

At four-loop order, we encounter integrals in the family
\begin{align}\label{eq:IntegralFamily}
  I^{\eta_{1}\dots\eta_{4}}_{n_{1}\dots n_{18}}=\int_{\ell_{1}\dots\ell_{4}}\frac{1}{D_{1}^{n_{1}}\dots D_{18}^{n_{18}}},
\end{align}
with integration measure
\begin{align*}
  \int_{\ell_{i}}=\tilde{\mu}^{2\varepsilon}\int\frac{{\rm d}^{d}\ell_{i}}{(2\pi)^{d}}2\pi\,\delta(v\cdot\ell_{i}-v\cdot k_{1}),
\end{align*}
depending on the regularization scale $\tilde\mu$ from \cref{eq:EH_action}; it is convenient to redefine this as $\tilde{\mu}^{2}=e^{\gamma_{\rm E}}\mu^{2}/4\pi$.
The propagators defining the family are
\begin{align*}
  D_{i=1,\dots,4}=\ell_{i}^{2}+\eta_{i}\iO,&\quad D_{i=5,\dots,8}=(\ell_{i-4}-k_{1})^{2}, \\
  D_{i=9,\dots,12}=(\ell_{i-8}-k_{2})^{2},&\quad D_{i=13,14,15}=(\ell_{1}-\ell_{i-11})^{2}, \\
  D_{i=16,17}=(\ell_{2}-\ell_{i-13})^{2},&\quad D_{18}=(\ell_{3}-\ell_{4})^{2}.
\end{align*}
A set of 30 master integrals $\vec{I}$ for this problem was found and evaluated in ref.~\cite{Brunello:2026lzf}, but we perform the calculation independently here. 

We solve our integrals using the method of canonical differential equations \cite{Henn:2013pwa}.
To begin, we differentiate our master integrals $\vec{I}$ with respect to the only scale in our integrals, $x=\sin(\theta/2)$.
IBP reducing these derivatives to relate them to $\vec{I}$ produces a set of a differential equations,
\begin{align}
  \frac{\rm d}{{\rm d}x}\vec{I}(x;\varepsilon)=M(x;\varepsilon)\vec{I}(x;\varepsilon).
\end{align}
We then seek a basis transformation $\vec{I} \to \vec{I}_{\rm c}$ such that $\varepsilon$ is factorized out:
\begin{align}
  \frac{\rm d}{{\rm d}x}\vec{I}_{\rm c}(x;\varepsilon)=\varepsilon M_{\rm c}(x)\vec{I}_{\rm c}(x;\varepsilon).
\end{align}
In this form, the system is solved by the path-ordered exponential
\begin{align}
  \vec{I}_{\rm c}(x;\varepsilon)=\mathcal{P}\exp\left[\varepsilon\int_{x_{0}}^{x}M_{\rm c}(x^{\prime}){\rm d}x^{\prime}\right]\vec{J}(\varepsilon),
\end{align}
where $x_0$ is the base point, and $\vec{J}(\varepsilon)$ is a vector of boundary constants at that base point; we use $x_{0}=0$.

The differential equations contain no new geometries at this loop order, with the same elliptic block found at three loops showing up here too \cite{Bautista:2026fcp}.
At $\varepsilon = 0$, this elliptic sector sastisfies the Picard-Fuchs equation
\begin{equation}\label{eq:picard_fuchs}
  \mathcal{L}(\varpi(x))=\left[\frac{d}{dx}\left(x(1-x^2)\frac{d}{dx}\right)-x\right]\varpi(x)=0.
\end{equation}
We find the transformation to $\varepsilon$-factorized form for the diagonal blocks using \texttt{CANONICA} and \texttt{INITIAL} \cite{Meyer:2017joq,Dlapa:2022wdu}; for the off-diagonal blocks we apply the algorithm described in \cite{BenjaminSauer:2026smu}.

In \cref{eq:IntegralFamily} we have indicated with the superscript the causality prescription for each propagator which may go on shell---i.e. active propagators.
As has been explained in ref.~\cite{Bautista:2026fcp}, we may select the amplitude, \cref{eq:Amplitude}, or the Magnusian, \cref{eq:MagnusAmplitude}, simply by altering the prescription we use.
Specifically, the amplitude is extracted by using a retarded prescription ($\eta_{i}=+1$) for all active propagators.
The Magnusian, on the other hand, is built from a certain weighted average of all possible prescriptions, which we dubbed the integral's Murua value in ref.~\cite{Bautista:2026fcp}.\footnote{This will have to be revisited when we consider subleading absorptive effects. We say more on this in the conclusion.}
A general form was identified in that reference for the appropriate weights: an integral with $n$ active propagators, of which $k$ have an advanced causality prescription $\eta_{i}=-1$, is given the weight
\begin{align*}
  \frac{1}{(n+1)!}{n\choose{k}}.
\end{align*}
This observation has since been proven in refs.~\cite{Brandhuber:2026dkd,Kim:2026wyv}, and its application to four active propagators reads
\begin{align}
  I^{(M;4)}&=\frac{1}{5}\left(I^{++++}+I^{----}\right) \\
  &+\frac{1}{20}\left(I^{+++-}+I^{++-+}+I^{+-++}+I^{-+++}\right)\notag \\
  &+\frac{1}{20}\left(I^{---+}+I^{--+-}+I^{-+--}+I^{+---}\right)\notag \\
  &+\frac{1}{30}\left(I^{++--}+I^{+-+-}+I^{-++-}\right)\notag \\
  &+\frac{1}{30}\left(I^{--++}+I^{-+-+}+I^{+--+}\right),\notag
\end{align}
Since the Murua value involves all possible prescriptions, it is applicable already at the level of boundary integrals.

What remains, then, is to determine the boundary integrals for every causality configuration.
It continues to be the case that the integrals we must solve possess only two regions in the boundary limit $x\rightarrow0$.
These are the general region, where all $\ell_{i}\sim1\gg x$ and in which integrals scale as $x^0$; and the forward region where $\ell_{i}\sim x$ for all loop momenta.
Integrals in the forward region depend on the scale as $x^{n-8\varepsilon}$ for $n\in\mathbb{Z}$, and where the $-8\varepsilon$ power comes from the integral measure.

Twelve boundary integrals were needed in the general region.
Several of them are either iteratively one-loop or factorize into products of lower loop integrals which we have solved in refs.~\cite{Bautista:2026qse,Bautista:2026fcp}.
We determine the remainder from high-precision numerical evaluations to at least 200 digits with \texttt{AMFlow} \cite{Liu:2022chg,Huang:2026rjb}, reconstructed on a basis of the transcendental numbers
\begin{align}
  \{\pi,\log2,\zeta_{3},{\rm Li}_{4}(1/2)\}.
\end{align}
We have followed ref.~\cite{Brunello:2026lzf} in the inclusion of the polylogarithm in this basis; reconstruction is not possible without it.
The reconstruction of numbers on this basis was validated through overdetermination tests and by verifying stability over a large range of fitted digits.

All ten  forward-region boundary integrals can be evaluated analytically through two strategies.
The first consists of constructing systems of equations via partial fraction relations and loop-by-loop evaluation of cuts \cite{Driesse:2024xad}, which can then be solved exactly only in terms of gamma functions.
The second is direct integration through Schwinger parameters, generally returning combinations of hypergeometric functions.
We expand these numerically in $\varepsilon$ to high precision, reconstructing the expansion coefficients as with the general-region integrals.

\section{Gravitational-wave scattering}\label{sec:Results}

\begin{figure*}
  \centering
  \includegraphics[width=\textwidth]{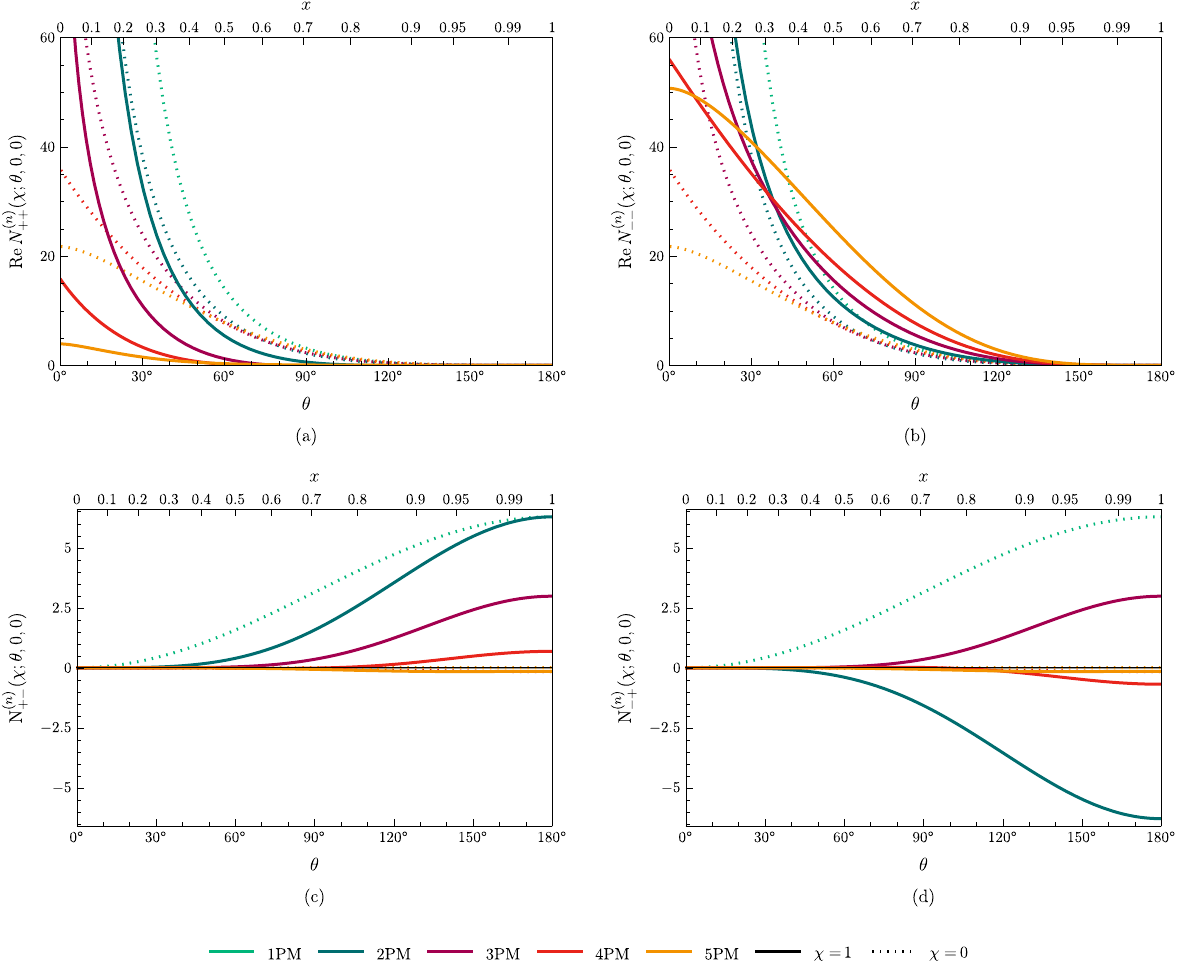}
  \caption{
    Gravitational-wave scattering for polar incidence ($\gamma=0$) in all four helicity configurations, for spinless ($\chi=0$) and extremal spin ($\chi=1$) scenarios.
    The top (bottom) two panels depict helicity-preserving(reversing) scattering.
    In the presence of spin, various helicity configurations are no longer related only by complex conjugation: the spin direction must be flipped as well.
    As such, anti-polar ($\gamma=\pi$) scattering is also encoded in these plots.
    Helicity-preserving scattering has imaginary, absorptive contributions.
    We only plot its real, conservative part.
  }
  \label{fig:NofTheta}
\end{figure*}

Combining with our previous work in refs.~\cite{Bautista:2026qse,Bautista:2026fcp}, we have produced every contribution to the amplitude and Magnusian listed in \cref{tab:AmplitudeTargets}.
Writing the latter according to \cref{eq:PMExpansion},
\begin{equation}
\begin{aligned}
  N_{h_{1},h_{2}}(\ePM&,\chi;\theta,\phi,\gamma) \\
  &=\sum_{n}\epsilon_{\rm PM}^{n}N_{h_{1},h_{2}}^{(n)}(\chi;\theta,\phi,\gamma)\,,
\end{aligned}
\end{equation}
defines the $n$PM Magnusian.
As mentioned above, it turns out that, up to the precision considered here, the dependence of the BHPT Magnusian on the spin parameter $\chi$ is polynomial, such that we may consistently additionally decompose our WQFT results as
\begin{align}
  N_{h_{1},h_{2}}^{(n)}(\chi;\theta,\phi,\gamma)=\sum_{s}\chi^{s}N_{h_{1},h_{2}}^{(n,s)}(\theta,\phi,\gamma)+\cO(G^6),
\end{align}
thus factorizing the dependence on both dimensionless quantities.
The functions $N^{(n,s)}_{h_{1},h_{2}}$ computed with the values in \cref{eq:BHPTMatchingCoeffs} for the Wilson coefficients are collected in the supplementary material at \zenodo.

Naturally, explicit expressions for the Magnusian with spin simplify dramatically for polar scattering, $\gamma=0$.
Our results in this configuration are illustrated in \cref{fig:NofTheta}.
As an example of the novel results we have produced, we print here the $\cO(\ePM^5\chi^1)$ Magnusian---whose contributions are summarized in \cref{tab:AmplitudeTargets}---for polar scattering.
We also set $\phi=0$ as it simply enters through an overall phase.
In terms of $x$, defined in \cref{eq:TransferMomentum}, this result is
\begin{widetext}
  \begin{equation}\label{eq:Magnusian51}
  \begin{aligned}
    &N^{(5,1)}_{1,-1}(\theta,0,0)=0 \\
    &N^{(5,1)}_{1,1}(\theta,0,0)=-\frac{4i\pi}{45}(1-x^2)^2-\frac{4\pi^{2}}{3}\frac{79 x^4+238 x^2+91}{1-x^2}G(-1;x) \\
    &+\left[1-\frac{12}{\pi^{2}}\,G(0,-1;x)\right]\frac{2\pi^{4}}{3}\frac{9+62x^{2}+57x^{4}+8x^{6}}{(x^{2}-1)^{2}}-\pi^{2}\frac{3375 x^6+81866 x^4+185911 x^2+56624}{384 \left(x^2-1\right)}K(x^{2})\mathcal{I}_{2}(x) \\
    &+\pi^2\frac{315 x^8+11835 x^7+343634 x^6+349394 x^5+3388003 x^4+583523 x^3+4384688 x^2+103344 x+668544}{5760 (1-x^2) (1+x)} \\
    &+\pi^{2}\frac{12995 x^6+204210 x^4+354075 x^2+84272}{384 \left(1-x^2\right)^2}\left[\frac{\mathcal{I}_{1}(x)}{K(x^{2})}-E(x^{2})\mathcal{I}_{2}(x)\right],
  \end{aligned}
  \end{equation}
\end{widetext}
and it is complex.
Its imaginary piece in the second line comes from $\mathcal{L}_{R^2,{\rm abs}}$ and maps on to the absorptive part of the BHPT result.
The remainder is real, coming from the rest of the action, and matches the BHPT phase shift.
As this example makes clear, the Magnusian provides an infrared-finite encoding of the scattering information also in the presence of spin.

All notation in \cref{eq:Magnusian51} is as in ref.~\cite{Bautista:2026fcp}.
$K(x^2)$ and $E(x^2)$ are complete elliptic integrals of the first and second kinds, and $\mathcal{I}_{1,2}(x)$ are the iterated integrals
\begin{equation}
\begin{aligned}
  \mathcal{I}_{1}(x)&\equiv\int_{0}^{x}{\rm d}y\,K(y^{2}), \\
  \mathcal{I}_{2}(x)&\equiv\frac{\pi}{2}-\int_{0}^{x}{\rm d}y\,\frac{\mathcal{I}_{1}(y)}{y(1-y^{2})K^{2}(y^{2})}.
\end{aligned}
\end{equation}
Finally, $G(a_{1},\dots,a_{n};x)$ are multiple polylogarithms.

Note that, for generic spin orientations, the Magnusian depends non-trivially on $\phi$.
Relatedly, it is complex even in the absence of time-asymmetric effects in such cases.

\subsection{Matching to BHPT}

We now elaborate on the partial-wave projections of the WQFT Magnusian, and report our findings from imposing the matching condition \cref{eq:Nprime-Delta-prime}.
The forward behavior of the Magnusian is improved in the presence of spin, such that the projection of the helicity-preserving result need not be regularized from $\cO(\ePM^2)$.

Now, a decomposition of the Magnusian is useful for performing the partial-wave projections analytically:
\begin{align}\label{eq:NEllipticDecomp}
  N(\ePM,\chi;\theta,\phi,\gamma)=N_{\rm n.e.}+N_{\rm e.},
\end{align}
where we have suppressed all helicity dependence and the functional dependence of the Magnusian components on the right-hand side.
In the first term on the right-hand side, the quantity $x$ enters either rationally or through multiple polylogarithms (MPLs).
With the help of \texttt{PolyLogTools} \cite{Duhr:2019tlz} for the MPLs, analytically projecting on a partial-wave basis is straightforward.


When the scattering reverses the helicity of the gravitational wave, the second term in \cref{eq:NEllipticDecomp} vanishes up to the precision considered.
For helicity-preserving scattering, this term depends on $x$ through elliptic functions, rendering analytic partial-wave projection more challenging.
Nevertheless, we developed a method in ref.~\cite{Bautista:2026fcp} which achieves an analytic projection of elliptic functions.
Let us review this method and explain the minor modifications needed to treat our results here.


Looking more closely at the elliptic sector of the Magnusian, we realize that we are able to write
\begin{align}
  N_{\rm e.}=A(x,\phi)q(x)+B(x,\phi)q'(x),
\end{align}
where $A(x,\phi)$ and $B(x,\phi)$ are rational functions of $x$, and depend further on the azimuthal angle $\phi$.
Additional dependence on variables inactive in the partial-wave transform, such as the BH spin parameter $\chi$ and the incidence angle $\gamma$, is implicit.
The function $q(x)$ captures all elliptic functions, and satisfies the Picard-Fuchs equation
\begin{equation}\label{eq:pf_q}
  \mathcal{L}\bigl(q(x)\bigr) = - f(x)\,,
\end{equation}
for the self-adjoint operator $\mathcal{L}$ defined in
\cref{eq:picard_fuchs}.
Crucially, $f(x)$ is free of elliptic functions in all cases we encounter.

\Cref{eq:N-projection} defines the partial-wave modes of the elliptic sector:
\begin{align}
  N_{\ell m,{\rm e.}}
  &= (-1)^{m-2}
  \int \mathrm{d}\Omega\,
  {}_{2}Y_{\ell,-m}(\theta_{\sigma},\phi)\,
  N_{\rm e.}
\nonumber\\
  &= \int_0^1 \mathrm{d}x\,
  \bigl[
    A_{\ell m}(x) q(x)
    + B_{\ell m}(x) q'(x)
  \bigr]\,.
\end{align}
Though the presence of spin complicates the $\phi$ dependence, integrating over this angle remains uncomplicated.
In passing to the second line, we have absorbed the factor $4x$ from the measure,
$\sin\theta\,\mathrm{d}\theta= 4x\,\mathrm{d}x$, as well as the $x$ dependence of the harmonics into the definitions of $A_{\ell m}(x)$ and $B_{\ell m}(x)$.
Quite conveniently, these functions turn out to be polynomial in $x$.

Next, we integrate the partial-wave projection by parts; in all cases we observe that boudary terms vanish (see ref.~\cite{Bautista:2026fcp} for details), such that we may write
\begin{equation}
\begin{aligned}
  N_{\ell m,{\rm e.}}
  &=
  \int_0^1 \mathrm{d}x\,
  R_{\ell m}(x) q(x)\,, \\
  R_{\ell m}(x)
  &\equiv
  A_{\ell m}(x) - B'_{\ell m}(x)\,.
\end{aligned}
\end{equation}
Consider now a polynomial $v_{\ell m}(x)$ whose image under $\mathcal{L}$ is $R_{\ell m}(x)$,
\begin{equation}
  \mathcal{L}\bigl(v_{\ell m}(x)\bigr)
  =
  R_{\ell m}(x)\,.
\end{equation}
Such a polynomial may be found by constructing an ansatz and imposing this relation.
Then, the self-adjointness of $\mathcal{L}$, the
Picard-Fuchs equation \cref{eq:pf_q}, and the regularity of $q(x)$ at the boundary points $x=0,1$ simplify the elliptic modes to
\begin{equation}\label{eq:EasyEllipticProjection}
  N_{\ell m,{\rm e.}}
  =
  - \int_0^1 \mathrm{d}x\,
  v_{\ell m}(x) f(x)\,.
\end{equation}
The remaining integration over $x$ does not involve elliptic functions, and can therefore be performed using standard methods.

Elliptic functions enter the Magnusian from three-loop order.
Here, the same functions $q(x)$ and $f(x)$, related through \cref{eq:pf_q}, appear in $N^{(4,0)}_{1,1}$ and $N^{(5,1)}_{1,1}$:
\begin{equation}
  q(x)=K(x^2)\mathcal{I}_2(x)\,,\quad \mathcal{L}\bigl(q(x)\bigr) = -1\,.
\end{equation}
That is, the same projection technique works in the presence or absence of spin.

At four loops, defining
\begin{equation}
\begin{aligned}
  &T_{r,a}(x)\\
  &=\int^{x}_{0}\frac{{\rm d}y}{y(y^{2}-1)K^{2}(y^{2})}\int^{y}_{0}{\rm d}z\frac{K(z^{2})}{z^{r}}G(a,0;z)
\end{aligned}
\end{equation}
allows us to write a compact expression for $q(x)$:
\begin{equation}
  q(x)=K(x^2)\bigg(\frac{\pi^3}{8}+T_{0,1}-T_{0,-1}+2T_{1,1}+2T_{1,-1}\bigg).
\end{equation}
This satisfies
\begin{align}\label{eq:pF_4loop}
  &\mathcal L(q(x)) \\
  &=-\left[\left(1+\frac2x\right)G(1,0;x)+\left(\frac2x-1\right)G(-1,0;x)\right].\notag
\end{align}
What is crucial in the above relation is that the right-hand side is indeed free of elliptics, facilitating the computation of \cref{eq:EasyEllipticProjection}.

With this integration method at hand, we are in a position to project all of our results onto a partial-wave basis analytically.
All components with $n<5$ immediately satisfy \cref{eq:Nprime-Delta-prime} upon partial wave transforming, thereby reproducing the BHPT result from bosonic-oscillator WQFT.
All projections were performed for $\ell\in[2,10]$ and $m\in[-\ell,\ell]$.

Imposing that \cref{eq:Nprime-Delta-prime} holds for $n=5$ uniquely fixes the free coefficients in \cref{eq:ConservativeR2,eq:AbsorptiveR2}.
Specifically, matching BHPT at $n=5$ and spin order $s$ imposes
\begin{alignat}{2}
  s=1\!:\quad&a_{(SE^2;0)}=\frac{16}{45}, &&a_{(SB^2;0)}=-\frac{4}{45},\notag \\
  s=2\!:\quad&c_{(S^2RE;0)}=-\frac{29}{280},\quad &&c_{(S^2B^2;0)}=\frac{1}{28},\notag \\
  &c_{(S^2E^2;0)}=-\frac{19}{126},\quad &&c_{(S^0B^2;2)}=-\frac{1}{840},\notag \\
  s=3\!:\quad&a_{(SE^2;2)}=\frac{16}{15}, &&a_{(SB^2;2)}=-\frac{3}{5},\label{eq:BHPTMatchingCoeffs} \\
  &a_{(S^3E^2;0)}=0, &&a_{(S^3B^2;0)}=-\frac{2}{3},\notag \\ 
  s=4\!:\quad&c_{(S^0R^2;4)}=-\frac{1}{48},\quad &&c^{\prime}_{(S^2B^2;2)}=-\frac{1}{8}.\notag
\end{alignat}
Since the operators multiplying the coefficients above are free of spatial derivatives of curvature tensors, they only have non-vanishing projections on, and their coefficients are fixed by, $\ell=2$ modes.
A strong check on our Magnusian is that it is gauge invariant, giving us confidence in the correctness of the matching, particularly at high loop/spin orders.

Translating the PM order to the loop order, the matching reported here at $s=2,4$ involves the two-loop and tree-level contributions to the amplitude from the $\cO(R)$ part of the action, respectively.
With the values from the $s=4$ matching we confirm the result of refs.~\cite{Bautista:2022wjf,Bautista:2023sdf} that the spin dependence of the tree-level amplitude exponentiates up to fourth order in the spin vector. Since the linear-in-curvature part of the action is time-reversal symmetric, the purely absorptive $s=1,3$ matching does not require any loop computations.
Additionally, we reproduce the result of ref.~\cite{Brunello:2026lzf} at $s=0$.

Let us remind the reader of all the pieces of the WQFT Lagrangian which describes Kerr BH dynamics up to $\cO(G^5)$:
\begin{equation}
\begin{aligned}
  \Lagr^{\rm Kerr}&=-\frac{m}{2}g_{\mu\nu}(x)\dot{x}^{\mu}(\tau)\dot{x}^{\nu}(\tau)-im\bar{\alpha}_{a}(\tau)\frac{\nabla\alpha^{a}(\tau)}{{\rm d}\tau} \\
  &+\mathcal{L}_{R}^{\rm Kerr}+\mathcal{L}_{R^2,{\rm cons}}^{\rm Kerr}+\mathcal{L}_{R^2,{\rm abs}}^{\rm Kerr}.
\end{aligned}
\end{equation}
Universal (i.e. body-independent) terms are written in the first line.
Non-minimal operators relevant for Kerr are written explicitly in \cref{eq:KerrLinearR,eq:ConservativeR2,eq:AbsorptiveR2}, and the values in \cref{eq:BHPTMatchingCoeffs} should be substituted for their Wilson coefficients.

\subsection{Change in mass and spin}\label{sec:Absorption}

As an immediate physical application of our action, we can compute the leading change in the mass and spin magnitude of a Kerr BH in a scattering event with a second compact body.
This is straightforward to do using WQFT.
For simplicity, we consider the second body to be spinless.
The action for this two-body system is, then,
\begin{align}
  S_{\rm two-body}
  =
  S_{\rm wl}
  +
  S_{\rm wl'}
  +
  S_{\rm EH}
  \ ,
\end{align}
where $S_{\rm wl}$ is given in eq.~\eqref{eq:GenAction} and
\begin{align}
  S_{\rm wl'}=-\frac{m'}{2}\int{\rm d}\tau'&g_{\mu\nu}(x')\dot{x}'^{\mu}(\tau)\dot{x}'^{\nu}(\tau)\,,
\end{align}
captures the second body's mass monopole.

The initial mass and spin magnitude for the primary body are given in terms of the tensorial parameters of the BH as
\begin{align}\label{eq:InitialMassSpin}
    m^2 = p^2
    \, ,\quad
    \chi^{2}=\frac{S_{0,\mu\nu}S_0^{\mu\nu}}{2G^{2}m^{4}}
    \, ,
\end{align}
with contractions in the initially flat spacetime.
Note that the initial momentum is related to the background quantities in \cref{eq:ExpandWL} through $p^\mu=mv^\mu$.
Given that $\chi$ is no longer a constant of the scattering, we define it in terms of the initial spin tensor, in contrast to \cref{eq:ChiParameter}.

Now, WQFT enables the direct calculation of the linear impulse and spin kick, in terms of which the change in mass and spin magnitude are
\begin{equation}\label{eq:changesInChiM}
\begin{aligned}
  &\Delta m^2
  =
  (p+\Delta p)^2-m^2
  \,,
  \\
  &\Delta (\chi^2 m^4) \\
  &\qquad=
  \frac{
    (S_{0,\mu\nu}+\Delta S_\mn)
    (S_0^{\mu\nu}+\Delta S^\mn)
  }{2G^2}
   -
  \chi^2m^4
  \,.
\end{aligned}
\end{equation}
Any contribution to $\Delta p^\mu$ or $\Delta S^\mn$ derived from interactions from the in-out action will lead to vanishing changes in $m$ and $\chi$.
Instead, contributions from the absorptive Lagrangian lead to non-zero absorptive changes.
Then, expanding eq.~\eqref{eq:changesInChiM} to leading order in absorptive effects gives
\begin{align}\label{eq:LOAbsorptiveObs}
  \Delta m
  &=
  v_\mu\Delta p^\mu_{\rm abs}
  +
  \dots
  \, ,
  \quad
  \Delta\chi
  =
  \frac{S_{0,\mu\nu}\Delta S^{\mu\nu}_{\rm abs}}{2G^2m^4\chi}
  +
  \dots
  \, .
\end{align}
Here we used that $\Delta m$ enters one power of $G$ later than $\Delta \chi$: for Kerr $\Delta m$ is $\mathcal{O}(G^6)$ \cite{Porto:2007qi,Goldberger:2020fot,Bautista:2024emt}, whereas $\Delta \chi$ starts at $\mathcal{O}(G^5)$ \cite{Goldberger:2020fot} (see also \cite{Poisson:2004cw,Saketh:2022xjb}).

\Cref{eq:LOAbsorptiveObs} indicates we need the longitudinal part of the absorptive impulse---as has also been observed in ref.~\cite{Jones:2023ugm}---and the spatial part of the absorptive spin kick.
Changes in tensorial parameters are given in the WQFT framework by one-point functions, and are made absorptive by using vertices from \cref{eq:AbsorptiveR2}:
\begin{align}\label{eq:absorptiongraph}
  \begin{tikzpicture}[baseline={([yshift=-.5ex]current bounding box.center)},scale=1.2]
  \def\xtop{.45}    
  \def\ytop{.15}    
  \def\outlen{.35}   
  \draw [dotted, thick] (-\xtop-\outlen,\ytop) -- (\xtop+\outlen,\ytop);
  \draw [dotted, thick] (-\xtop-\outlen,-0.6) -- (0,-0.6);
  \draw [zParticle] (0,-0.6) -- (\xtop+\outlen,-0.6);
  \draw [photonTest] (-\xtop,\ytop) -- (0,-0.6);
  \draw [example1={.55}] (-\xtop,\ytop) -- (0,-0.6);
  \draw [photonTest] (\xtop,\ytop) -- (0,-0.6);
  \draw [example1={.55}] (\xtop,\ytop) -- (0,-0.6);
  \draw [fill=black] (-\xtop,\ytop) circle (.067);
  \draw [fill=black] (\xtop,\ytop) circle (.067);
  \draw [fill=black] (-0.07,-0.67) rectangle (0.07,-.53);
  \end{tikzpicture}\,\, .
\end{align}
Choosing the external leg to be $\delta x^\mu$ or $\delta\alpha^\mu$ and setting its energy to zero, this graph is then proportional to $\Delta p^\mu_{\rm abs}$ or $\Delta\alpha^\mu_{\rm abs}$, respectively.
The absorptive spin kick is composed of the latter and its conjugate, $\Delta \bar\alpha_{\rm abs}^\mu$.
However, in this leading-order calculation, we find it simpler to use an ``outgoing'' $S^\mn$ arriving directly at $\Delta S^\mn$.
Details are given in \Cref{app:absComputations}, but this makes use of the background Poisson bracket of $S_0^{\mn}$ with the local frame \cite{Kim:2026bqi},
\begin{align}\label{eq:SpinFrameBracket}
  \{S_0^{\mu\nu},\Lambda^\rho_{0A}\}
  =
  -\eta^{\rho\mu}\Lambda^\nu_{0A}
  +
  \eta^{\rho\nu}\Lambda^\mu_{0A}
  \, .
\end{align}
In this equation, $S_0^\mn$ generates rotations of $\Lambda^\mu_{0A}$, the initial (background) value of $\Lambda^\mu_A(\tau)$.

Aside from accommodating the dynamics of $\Lambda^\mu_A$, discussed in Appendix~\ref{app:absComputations}, evaluating the graphs~\eqref{eq:absorptiongraph} involves quite standard one-loop techniques (see, e.g., ref.~\cite{Jakobsen:2021zvh}).
Carrying out the computation, we find
\begin{align}\label{eq:changeinmass}
  \Delta m
  &=
  \frac{\pi G^6 m^5 m'{}^2\, \chi\, n_L\, \gamma}{8\, |b|^6}
  \Big[\,
  3{-}7\gamma^2
  {+}\frac{3\chi^2}{8}
  \Big(
  99-75n_b^2
  \notag\\
  &\qquad
  -110n_L^2+21\big(5n_b^2+5n_L^2-6\big)\gamma^2
  \Big)
  \Big]
  +
  \mathcal{O}(G^7)
  \ ,
\end{align}
and
\begin{align}\label{eq:changeinspin}
  \Delta\chi
  =
  \frac{\pi G^5 m^3 m'{}^2\, \chi}{40\, |b|^5 \sqrt{\gamma^2-1}}
  \Big[\,
  P_1-\frac{3\chi^2}{8}P_3
  \,\Big]
  +\mathcal{O}(G^6)
  \ ,
\end{align}
with
\begin{equation}
\begin{aligned}
  P_1
  &=
  -15+11n_b^2+16n_L^2 
  -5\big(20n_b^2+25n_L^2-27\big)\gamma^2
  \\
  &\qquad
  +35\big(3n_b^2+3n_L^2-4\big)\gamma^4
  \ ,
  \\
  P_3
  &=
  145-128n_b^2+15n_b^4 
  +16n_{L}^{2}(5n_{b}^{2}+15n_{L}^{2}-23)
  \\
  &\quad
  {-}10\big(113+15n_b^4{+}40n_L^2(n_L^2-4)
  {+}5n_b^2(11n_L^2{-}23)\big)\gamma^2
  \\
  &\quad
  +35\big(32+5n_b^4-34n_L^2+5n_L^4 
  +2n_b^2(5n_L^2-17)\big)\gamma^4
  \ .
\end{aligned}
\end{equation}
We are abusing notation slightly here: rather than the spin vector's polar angle, $\gamma=v\cdot v'$ in these results, where $v'{}^{\mu}$ is the four velocity of the second compact object.

Other than this scalar, the results depend on the impact parameter $b^\mu$ with magnitude $|b|^2=-b^\mu b_\mu$, and on the initial dimensionless spin $\chi$ and ring radius vector through
\begin{align}
  n_L = -\frac{a\cdot L}{|a||L|}
  \ ,
  \qquad
  n_b = -\frac{a\cdot b}{|a||b|}
  \ .
\end{align}
Recall that $|a|^2=-a^\mu a_\mu$.
As $b^\mu$ enters only quadratically, its sign doesn't play a role in either observable.
However, letting it point from $m$ to $m'$, the unit-normalized angular momentum vector is
\begin{align}
  \frac{L^\mu}{|L|}
  =
  -\frac{\epsilon^\mu{}_{\nu\ab}
  b^\nu v^\alpha v'{}^\beta}{\sqrt{\gamma^2-1}|b|}
  \, ,
\end{align}
where, again, $|L|^2=-L^\mu L_\mu$.

Our result for the change in mass, \cref{eq:changeinmass} agrees with that reported in eq.~(83) of ref.~\cite{Bautista:2024emt} for generic spin orientations.
To the best of our knowledge, the result for the change in spin, \cref{eq:changeinspin}, is new.\footnote{Refs.~\cite{Goldberger:2020fot,Saketh:2022xjb} (see also \cite{LeTiec:2020bos}) present explicit  equations
for the rate of mass and spin
dissipation of BHs coupled to tidal backgrounds, including PN binary systems. 
We expect that time integration of these quantities along a scattering trajectory will overlap with the small-velocity expansion of our result.}

As another cross check on our method, we have carried out the same computation at leading order for the change in mass and angular momentum of a Schwarzschild BH scattering off a secondary spinless body.
These quantities are controlled at leading order by the absorptive coupling
\begin{align}
  E_{ab}^{(-)}
  \frac{\mathrm{d}}{\mathrm{d}\tau}
  E^{(+)ab}
  \ ,
\end{align}
whose Wilson coefficient is easily obtained by matching to the leading-in-$\ePM$ absorptive amplitude for gravitational-wave/Schwarzschild scattering of ref.~\cite{Bautista:2024emt}. 
We find complete agreement with the leading-order results reported for the first time in ref.~\cite{Goldberger:2020wbx} for the change in mass, and that in ref.~\cite{Cipriani:2026myb} for the change in angular momentum.



\section{Conclusion}

We have augmented the matching of WQFT to BHPT developed in refs.~\cite{Bautista:2026qse,Ivanov:2026icp,Bjerrum-Bohr:2026fhs} to incorporate spin, with the bosonic-oscillator formulation of ref.~\cite{Haddad:2024ebn} as the underlying effective framework.
Specifically, we have pinpointed a WQFT action which produces the entire BHPT amplitude---including both conservative and absorptive effects---for gravitational-wave scattering up to $\cO(G^5)$, in a counting scheme where powers of spin scale linearly with Newton's constant.

Naturally, the inclusion of spin leads to computational and conceptual complications.
First, care must be taken in the matching to BHPT in ensuring the comparison is done for modes defined in the same frames.
At the core of this is that the convenient frame where the gravitational wave impinges on the BH from along the $z$-axis is not the frame in which BHPT results are most easily extracted.
This is solved by a simple rotation of the modes from one frame to the other.
Second, we showed how the introduction of an effective two-graviton vertex allows one to avoid a proliferation of graphs by recycling those relevant to the spinless case.

In performing the matching at two loops and quadratic order in spin, we uncovered the need for new $\cO(S^2R^2)$ operators.
As BHs are known to possess vanishing static Love numbers, it is unclear how these operators correspond to properties of a Kerr BH.
Thus, though we have identified an action which faithfully produces Kerr BH dynamics, a deeper understanding of the relation between the degrees of freedom of a BH and of the EFT remains to be explored. 

Nevertheless, these matched operators have immediate implications for PM calculations of binary BH scattering, where they become relevant at quadratic-in-spin and three-loop orders.
At the moment, the three-loop spin frontier lies at linear order in spin \cite{Jakobsen:2023hig}.
Surpassing this frontier requires the inclusion of these new operators in order to accurately describe Kerr BH scattering.

Completely matching to the BHPT amplitude at $\cO(G^5)$ meant also reproducing its absorptive part which first enters at this order.
We resorted to the in-in formalism to introduce the necessary time-asymmetric operators.
With their coefficients fixed by the matching, we could with little effort study the leading change in mass and spin of a Kerr BH in a scattering encounter with a spinless compact object.
While the former is known, the latter is a novel result.

Future analyses involving loop-level absorptive contributions will require us to revisit and potentially modify the matching procedure employed here.
The reason for this is that the in-in formalism imposes a definite flow of causality through time-asymmetric Feynman vertices, producing a tension with the construction of Murua values of integrals.
A reliable fall-back is to match the absorption coefficient at the level of the amplitude and the phase shift at the level of the Magnusian, but an efficient prescription or unified approach remains to be explored.

Finally, we are now one PM order away from resolving the question of the effective description of a Kerr BH above the fourth order in its spin tensor.
Our immediate next step is to undertake the evaluation of the amplitude for a gravitational-wave scattering off of a Kerr BH at $\cO(G^6)$.
In this pursuit, it would be interesting to additionally encode our action in irreducible $SO(d-1)$ operators in order to facilitate a physical interpretation of the Wilson coefficients which appear.

\acknowledgements

We are grateful to Giacomo Brunello, Mario Meo, and Sid Smith for helpful discussions and collaboration on related work.
We also thank Jung-Wook Kim, Gustav Mogull, Raj Patil, Julio Parra-Martinez,  Johannes Pirsch, Trevor Scheopner, and Mao Zeng for helpful discussions, and Dogan Akpinar for compiling and sharing results from ref.~\cite{Akpinar:2025byi} for comparison.
This work has made use of the Black Hole Perturbation Toolkit \cite{BHPToolkit} and \texttt{FeynCalc} \cite{Mertig:1990an,Shtabovenko:2016sxi,Shtabovenko:2020gxv,Shtabovenko:2023idz}.
Integrand generation and tensor reduction were done using \texttt{FORM} \cite{Davies:2026cci} and \texttt{OPITeR} \cite{Goode:2024cfy}. IBP reduction was performed with \texttt{Kira 3} \cite{Lange:2025fba}.
Canonicalization was performed with the help of \texttt{CANONICA} \cite{Meyer:2017joq} and \texttt{INITIAL} \cite{Dlapa:2022wdu}.
Numerical integration was performed with \texttt{AMFlow} \cite{Liu:2022chg,Huang:2026rjb}.
The work of M.D. and K.H.~was funded by the European Union through the 
European Research Council under ERC Advanced Grant 101097219 (GraWFTy). The work of
Y.F.B. was funded by the European Union (ERC, Classical$+$, 101200505).
Views and opinions expressed are however those of the authors only and do not necessarily reflect those of the European Union or European Research Council Executive Agency. Neither the European Union nor the granting authority can be held responsible for them.

\subsection*{Statement on AI usage}

Claude Fable 5.1 and Opus 5.5 were used to optimize code to handle large expressions; bug hunting; aid in computing absorptive observables; generating plots of the results; and aid in proofreading.
GPT-6 Astra was used to recast the elliptic part of the 4-loop amplitude in terms of the $q(x)$ function.
All output was human reviewed, reproduced, checked numerically, and/or validated against earlier results.
All writing and scientific ideas in this paper are human.

\section*{Data availability}

All results for the scattering amplitude and Magnusian computed in this work are available at \zenodo.
There we also compile our results from refs.~\cite{Bautista:2026qse,Bautista:2026fcp}, and the BHPT phase shift and absorption factor expanded to commensurate orders as our amplitudes.

\appendix

\section{Proof that the spin Casimir is non-dynamical}\label{app:ChiConst}

Consider an operator proportional to the spin Casimir $\chi^2$ defined in \cref{eq:ChiParameter},
\begin{align}
  \mathcal{O}=S_{\alpha\beta}S^{\alpha\beta}\mathcal{O}_{1},
\end{align}
where $\mathcal{O}_{1}$ is arbitrary.
We wish to prove that any variation of $S_{\alpha\beta}S^{\alpha\beta}$ has a vanishing contribution to the gravitational-wave scattering amplitude computed in the main body of the paper.

Towards this end, note that the only perturbative degree of freedom on which the Casimir depends is the bosonic oscillator,
\begin{align}
  S_{\alpha\beta}S^{\alpha\beta}=S_{ab}S^{ab}=-2im\bar\alpha_a\alpha_bS^{ab},
\end{align}
having used \cref{eq:Vielbein} to convert curved- to flat-space indices.
Thus the Casimir is trivially inert under any variation of $\cO$ with respect to the worldline or metric perturbations, as well as oscillator perturbations which act only on $\cO_{1}$.

All diagrams contributing to gravitational-wave/compact-object scattering involve vertices with at most one oscillator perturbation, so what is left to analyze is how
\begin{equation}\label{eq:VaryCasimir}
\begin{aligned}
  \cO_{1}\frac{\partial}{\partial\alpha_{c}}S_{\alpha\beta}S^{\alpha\beta}&=-4im\cO_{1}\bar\alpha_aS^{ac}, \\
  \cO_{1}\frac{\partial}{\partial\bar\alpha_{c}}S_{\alpha\beta}S^{\alpha\beta}&=4im\cO_{1}\alpha_aS^{ac}
\end{aligned}
\end{equation}
may enter amplitudes.
Consider then a second arbitrary operator $\cO_{2}$.
Its variations with respect to an oscillator are
\begin{equation}\label{eq:VaryO2}
\begin{aligned}
  \frac{\partial\cO_{2}}{\partial\alpha_{c}}&=\frac{\partial S^{ab}}{\partial\alpha_{c}}\frac{\partial\cO_{2}}{\partial S^{ab}}=-2im\bar{\alpha}^{a}\frac{\partial\cO_{2}}{\partial {S^{a}}_{c}} \\
  \frac{\partial\cO_{2}}{\partial\bar\alpha_{c}}&=\frac{\partial S^{ab}}{\partial\bar\alpha_{c}}\frac{\partial\cO_{2}}{\partial S^{ab}}=2im\alpha^{a}\frac{\partial\cO_{2}}{\partial {S^{a}}_{c}},
\end{aligned}
\end{equation}
where we have used that the derivative with respect to $S^{ab}$ is antisymmetric in its indices.
We may now assess the contributions of \cref{eq:VaryCasimir,eq:VaryO2} to pairs of diagrams with opposite oscillator propagators:
\begin{align}\label{eq:SumOscDiagrams}
  &\begin{tikzpicture}[baseline=-8mm,scale=1.2]
    \draw [zParticle, thick] (-.5, -0.6) -- (.5, -0.6);
    \draw [fill = white] (-0.5,-0.6) circle (.2);
    \draw [pattern = north east lines] (-0.5,-0.6) circle (.2)  node [left] {$\displaystyle \cO_{1}\frac{\partial S^2}{\partial\alpha_{c}}\quad$};
    \draw [fill = white] (0.5,-0.6) circle (.2);
    \draw [pattern = north east lines] (0.5,-0.6) circle (.2) node [right] {$\displaystyle \quad\frac{\partial\cO_{2}}{\partial\bar\alpha_{d}}$};
  \end{tikzpicture} \\
  &+
  \begin{tikzpicture}[baseline=-8mm,scale=1.2]
    \draw [zParticle, thick] (-.5, -0.6) -- (.5, -0.6);
    \draw [fill = white] (-0.5,-0.6) circle (.2);
    \draw [pattern = north east lines] (-0.5,-0.6) circle (.2)  node [left] {$\displaystyle \cO_{1}\frac{\partial S^2}{\partial\bar\alpha_{c}}\quad$};
    \draw [fill = white] (0.5,-0.6) circle (.2);
    \draw [pattern = north east lines] (0.5,-0.6) circle (.2) node [right] {$\displaystyle \quad\frac{\partial\cO_{2}}{\partial\alpha_{d}}$};
  \end{tikzpicture}=-8im\cO_{1}\cY S_{ab}\frac{\partial\cO_{2}}{\partial {S_{b}}^{c}}S^{ca}.\notag
\end{align}
Here we need not concern ourselves with the rest of the expression encoded in the blobs, beyond that it is common to both diagrams and represented by $\cY$.
Due to anti-symmetry of the three factors involving the spin tensor, this sum of diagrams vanishes:
\begin{align}
  S_{ab}\frac{\partial\cO_{2}}{\partial {S_{b}}^{c}}S^{ca}={\rm Tr}\left(S\cdot\frac{\partial\cO_{2}}{\partial S}\cdot S\right)=0.
\end{align}
Diagrams with oscillator propagators must always appear in such pairs in order to reconstruct a spin tensor.
Thus, we conclude that the spin Casimir is truly non-dynamical in our computations of the amplitude, so we are justified in treating it as a constant in our action.

\section{Proof of vanishing recoil}\label{app:RecoilProof}

As observed in ref.~\cite{Brunello:2026rdk} and proven in ref.~\cite{Correia:2026utp} for spinless scattering, all recoil contributions vanish in the transverse-traceless (TT) gauge, i.e. when the quantities $v \cdot \epsilon_i$ vanish.
Here we provide a simple, all-order argument based on WQFT diagrammatics that there is no recoil in this gauge up to $\cO(S^1)$, and no spin recoil up to $\cO(S^2)$.

In the WQFT language, recoil is indicated by the presence of at least one trajectory propagator in a diagram.
Consider an arbitrary such diagram:
\begin{align}\label{eq:RecoilDiagram}
  \begin{tikzpicture}[baseline=-5mm,scale=1.2]
    \draw [dotted, thick] (-.85, -0.6) -- (-.5, -0.6);
    \draw [dotted, thick] (.5, -0.6) -- (.85, -0.6);
    \draw [zParticle, thick] (-.5, -0.6) -- node [midway, above] {$\delta x$} (.5, -0.6);
    \draw [photonTest] (.5,-0.6) -- (.75,0.) node [right] {$k_{2}$};
    \draw [example1={.7}] (.5,-0.6) -- (0.75,-0.);
    \draw [photonTest] (-.5,-0.6) -- (-.75,0.) node [left] {$k_{1}$};
    \draw [example1={.55}] (-0.75,-0.) -- (-.5,-0.6);
    \draw [fill = white] (-0.5,-0.6) circle (.2);
    \draw [pattern = north east lines] (-0.5,-0.6) circle (.2);
    \draw [fill = white] (0.5,-0.6) circle (.2);
    \draw [pattern = north east lines] (0.5,-0.6) circle (.2);
    \node at (-0.5,-1.1) {$L^{\mu}$};
    \node at (0.5,-1.1) {$R^{\nu}$};
  \end{tikzpicture}.
\end{align}
This diagram factorizes into a left current $L^\mu$, which can only contain $k_{1}^{\mu}$, $\epsilon_{1}^{\mu}$, $v^\mu$, and $S_{0}^{\mu\nu}$; and a right current $R^{\mu}$ which is a function of $k_{2}^{\mu}$, $\epsilon_{2}^{\mu}$, $v^\mu$, and $S_{0}^{\mu\nu}$.
We make no assumptions about the loop order of either current.
Having absorbed polarizations into the currents, both possess only one index because they must form a scalar with the numerator of the worldline propagator, namely $L_{\mu}\eta^{\mu\nu}R_{\nu}$.

In particular, the single free index on each factor means that at most one polarization vector can cross the trajectory propagator to the other side of the diagram.
In this case, the left factor must contain at least one of either $k_{1}\cdot\epsilon_{1}=0$ (transversality), $v\cdot\epsilon_{1}=0$ (TT gauge), or $(\epsilon_{1}\cdot S_{0})^{\mu}\neq0$.
Similarly, the only non-vanishing possibility in the right factor is $(\epsilon_{2}\cdot S_{0})^{\mu}$.
If no polarization crosses the propagator then the only additional possible scalar product in either factor is $\epsilon_{i}^{2}=0$.
We conclude that in order for \cref{eq:RecoilDiagram} to survive, each factor must be at least $\cO(S^1)$, meaning the diagram is at least $\cO(S^2)$.

When the trajectory propagator is replaced by an oscillator propagator, it appears a polarization vector trapped on its side of a diagram may contract with a lone oscillator, $\epsilon_{i}\cdot\alpha_0$ or $\epsilon_{i}\cdot\bar\alpha_0$.
To see what happens in this case, note that both the open index on, say, $L^c$, and the index on the lone oscillator emerge after a derivative with respect to the spin tensor, as in the right-hand sides of \cref{eq:VaryO2}.
This means that
\[
  L^c\supset (\epsilon_{1}\cdot\alpha_0)\epsilon_{1}^{c}
\]
would come from the off-shell $S^{ac}h_{ac}=0$, vanishing by the opposite symmetries of the spin and graviton tensors.
The same holds if $\alpha_0\rightarrow\bar\alpha_0$ or $L^c\rightarrow R^c$ above.

The upshot is that, even in the presence of an unpaired oscillator, the only viable contraction in TT gauge is $(\epsilon_{i}\cdot S_{0})^c$ in each of $L^c$ and $R^c$.
A companion diagram with the conjugate oscillator propagator (i.e. as in \cref{eq:SumOscDiagrams}) turns the bare oscillators into an additional spin tensor, meaning a surviving diagram with oscillator propagator must be at least $\cO(S^3)$.


In explicit computations, this vanishing does not manifest itself until after IBP reduction.
The reason for this is that our tensor reduction method projects loop momenta onto components parallel and orthogonal to the 3-dimensional space spanned by $v^\mu$, $k_1^\mu$, and $k_2^\mu$.
This can introduce spurious dependence of $L_\mu$ or $R_\mu$ on the graviton momentum of which it is not a function.

\section{One-loop computations of absorption}\label{app:absComputations}

The leading-order diagram describing absorption has no internal propagation of worldline fields (except, in principle, the quadrupole fields which have been integrated out).
Therefore, the only propagating fields inside of the diagram are gravitons.
It is then convenient, in our case, to derive the two-body observables using the WQFT in-in free energy~\cite{Jakobsen:2022psy}.
This approach simply instructs us to compute the relevant WQFT vacuum diagrams keeping the minus Keldysh variables unconstrained---that is, non-zero.
One should, however, still keep only terms linear in these variables.
Further, following the in-in Feynman rules, all retarded propagators point toward the insertion of the minus field~\cite{Jakobsen:2022psy}.

The computational goal is therefore
\begin{align}
  iN_{\rm abs}
  \ =\ 
  \frac12
  \ \ 
  \begin{tikzpicture}[baseline={([yshift=-.5ex]current bounding box.center)},scale=1.2]
  \def\xtop{.45}    
  \def\ytop{.15}    
  \def\outlen{.35}   
  \draw [dotted, thick] (-\xtop-\outlen,\ytop) -- (\xtop+\outlen,\ytop);
  \draw [dotted, thick] (-\xtop-\outlen,-0.6) -- (0,-0.6);
  \draw [dotted, thick] (0,-0.6) -- (\xtop+\outlen,-0.6);
  \draw [photonTest] (-\xtop,\ytop) -- (0,-0.6);
  \draw [example1={.55}] (-\xtop,\ytop) -- (0,-0.6);
  \draw [photonTest] (\xtop,\ytop) -- (0,-0.6);
  \draw [example1={.55}] (\xtop,\ytop) -- (0,-0.6);
  \draw [fill=black] (-\xtop,\ytop) circle (.067);
  \draw [fill=black] (\xtop,\ytop) circle (.067);
  \draw [fill=black] (-0.07,-0.67) rectangle (0.07,-.53);
  \end{tikzpicture}
  \ .
\end{align}
Retarded physical observables are then given by Poisson brackets of this free energy with the background variables:
\begin{equation}
\begin{aligned}
  \Delta p_{\rm abs}^{\mu}
  &=
  \{ N_{\rm abs},p^{(+)\mu}\}
  \\
  \Delta S_{\rm abs}^\mn
  &=
  \{ N_{\rm abs},S^{(+)\mn}_{0}\}\,.
\end{aligned}
\end{equation}
The main benefit is that we construct the on shell outgoing line of the WQFT diagram via Poisson brackets of the background variables~\cite{Haddad:2024ebn}.
This sidesteps the need for a derivation of the propagator and dynamics of the body-fixed frame $\Lambda^{\mu}_{A}$.

Let us discuss the required Poisson brackets.
The in-in action in the (1)-(2) basis reads,
\begin{align}
  S_{\rm in-in}
  =
  S^{(1)}[\mathcal{O}^{(1)}]
  -
  S^{(2)}[\mathcal{O}^{(2)}]
  \, .
\end{align}
Here, $S^{(1)}$ ($S^{(2)}$) is the in-out action in terms of fields ($\mathcal{O}^{(i)}$ with $\mathcal{O}\in\{x^\mu,S^\mn,\Lambda^\mu_A,g_\mn\}$) moving forward (backward) in time.
For the free fields, the spinning action in terms of the frame (i.e. the ``relativistic spherical top'')~\cite{Porto:2016pyg,Ben-Shahar:2023djm,Haddad:2024ebn,Kim:2026bqi} reads,
\begin{align}
  S^{(i)}
  \!=\!
  \int \mathrm{d}\tau
  \Big(
    \!-
    \frac{m}2 
    \eta_\mn
    \dot x^{(i)\mu}
    \dot x^{(i)\nu}
    \!-\!
    \frac12 
    \eta_\mn
    \dot \Lambda_{A}^{(i)\mu }
    \Lambda_{B}^{(i)\nu }
    S^{(i)AB}
  \Big)
  \, .
\end{align}
The Poisson brackets of the variables of this action are provided in ref.~\cite{Kim:2026bqi}.
Particularly, for the Poisson bracket of the body-fixed frame with the spin tensor moving forward in time, one finds
\begin{align}\label{eq:generalBrackets}
  \{S^{(1)\mu\nu},\Lambda_{{A}}^{(1)\rho}\}
  =
  -\eta^{\rho\mu}\Lambda_{{A}}^{(1)\nu}
  +
  \eta^{\rho\nu}\Lambda_{{A}}^{(1)\mu}
  \, .
\end{align}
The brackets for the $(2)$ fields moving backward in time pick up a relative sign because their kinetic term has a relative sign compared with the $(1)$ fields.

One may straightforwardly specialize eq.~\eqref{eq:generalBrackets} to the initial-state background variables (as in e.g. ref.~\cite{Haddad:2024ebn}).
We find,
\begin{align}
  \{S^{(i)\mu\nu}_{0}
  ,
  \Lambda_{0{A}}^{(j)\rho}\}
  =
  \delta^{ij}
  \text{sgn}_i
  \Big(
  -\eta^{\rho\mu}\Lambda_{0{A}}^{(i)\nu}
  +
  \eta^{\rho\nu}\Lambda_{0{A}}^{(i)\mu}
  \Big)
  \, ,
\end{align}
with $\text{sgn}_1=1$ and $\text{sgn}_2=-1$ and $\Lambda^{(i)\mu}_{0A}$ the background value of $\Lambda^{(i)\mu}_{A}(\tau)$.
For completeness, in our conventions, the remainging Poisson brackets read,
\begin{align}
  \{b^{(i)\mu},p^{(j)\nu}\}
  &=
  -\delta^{ij}\text{sgn}_{i}
  \eta^\mn
  \ ,
  \\
  \{
    S^{(i)\mn}_0,
    S^{(j)\alpha\beta}_0
  \}
  &=
  \delta^{ij}\text{sgn}_i
  \big(
    \eta^{\mu\alpha}
    S_0^{(i)\nu\beta}
    +
    \eta^{\nu\beta}
    S_0^{(i)\mu\alpha}
    \\
    &\quad
    -
    \eta^{\mu\beta}
    S_0^{(i)\nu\alpha}
    -
    \eta^{\nu\alpha}
    S_0^{(i)\mu\beta}
  \big)
  \ .
  \notag
\end{align}
We note that since the (1)-(2) fields don't interact (except through the boundary conditions at future infinity), their Poisson brackets do not mix.
For $i=j=1$, these agree with refs.~\cite{Haddad:2024ebn,Haddad:2025cmw}.

It is now a straightforward matter to switch to the (+)/(-) Keldysh basis defined from the (1)-(2) basis via,
\begin{equation}\label{eq:plusminusdef}
\begin{aligned}
  \mathcal{O}^{(+)}
  &=
  \frac12
  \big(
    \mathcal{O}^{(1)}
  +\mathcal{O}^{(2)}
  \big)
  \ ,
  \\
  \mathcal{O}^{(-)}
  &=
  \mathcal{O}^{(1)}
  -
  \mathcal{O}^{(2)}
  \ .
\end{aligned}
\end{equation}
In particular, one finds the following brackets:
\begin{equation}\label{eq:PoissonBracketsPM}
\begin{aligned}
  \{
    b^{(-)\mu}
    ,
    p^{(+)\nu}
  \}
  &=
  -\eta^\mn
  \ ,
  \\
  \{
    \Lambda^{(-)\sigma}_A
    ,
    S^{(+)\mn}
  \}
  &=
  \eta^{\sigma\mu}\Lambda^{(+)\nu}_A
  -
  \eta^{\sigma\nu}\Lambda^{(+)\mu}_A
  \ .
\end{aligned}
\end{equation}
A similar results holds for the spin tensor bracket but we will not need it.
Brackets of plus fields with plus fields or minus fields with minus fields are proportional to minus fields and therefore vanish weakly (i.e. when putting minus fields to zero).

The required vertex rule is,
\begin{align}\label{eq:fRulesPP}
  &\begin{tikzpicture}[baseline=-5mm,scale=1.2]
  \draw [dotted, thick] (-.75,-0.6) -- (.75,-0.6);
  \draw [photonTest] (-.75,0.) -- (0,-0.6);
  \draw [example1={.55}] (-.75,0.) -- (0,-0.6);
  \draw [photonTest] (.75,0.) -- (0,-0.6);
  \draw [example1={.55}] (.75,0.) -- (0,-0.6);
  \draw [fill=black] (-0.07,-0.67) rectangle (0.07,-.53);
  \node [above left,  font=\footnotesize]  at (-.74,-.1) {$\mu_1\nu_1$};
  \node [above right,  font=\footnotesize] at (.74,-.1)  {$\mu_2\nu_2$};
  \path (0,-0.6) -- (-.75,0) node [pos=.75, below left,  font=\footnotesize, inner sep=2pt] {$k_1$};
  \path (0,-0.6) -- (.75,0)  node [pos=.75, below right, font=\footnotesize, inner sep=2pt] {$k_2$};
  \end{tikzpicture}
  =
  \frac{i\kappa_d^2}{4}
  V^{\mu_1\nu_1\mu_2\nu_2}(k_1,k_2)
  \, .
\end{align}
As mentioned, minus variables should be kept non-zero.
The fact that the causal flow points into this vertex then implies that it is linear in the minus variables, particularly $b^{(-)\mu}$, $v^{(-)\mu}$ and $\Lambda^{(-)\mu}_{0A}$. 
If the graviton would be endowed with an initial value as in refs.~\cite{Bautista:2026qse,Kim:2025hpn}, then it, too, would appear here.

As a simple, concrete example, the Poisson bracket of $p^{(+)\mu}$ with the Feynman vertex~\eqref{eq:fRulesPP} gives the on-shell vertex of an outgoing worldline deflection:
\begin{align}
  &\begin{tikzpicture}[baseline=-5mm,scale=1.2]
  \draw [dotted, thick] (-.75,-0.6) -- (0,-0.6);
  \draw [zParticle] (0,-0.6) -- (.75,-0.6);
  \node [below, font=\footnotesize] at (.6,-0.6) {$\rho$};
  \node [right, font=\footnotesize] at (.75,-0.6) {$\omega=0$};
  \draw [photonTest] (-.75,0.) -- (0,-0.6);
  \draw [example1={.55}] (-.75,0.) -- (0,-0.6);
  \draw [photonTest] (.75,0.) -- (0,-0.6);
  \draw [example1={.55}] (.75,0.) -- (0,-0.6);
  \draw [fill=black] (-0.07,-0.67) rectangle (0.07,-.53);
  \node [above left,  font=\footnotesize]  at (-.74,-.1) {$\mu_1\nu_1$};
  \node [above right,  font=\footnotesize] at (.74,-.1)  {$\mu_2\nu_2$};
  \path (0,-0.6) -- (-.75,0) node [pos=.75, below left,  font=\footnotesize, inner sep=2pt] {$k_1$};
  \path (0,-0.6) -- (.75,0)  node [pos=.75, below right, font=\footnotesize, inner sep=2pt] {$k_2$};
  \end{tikzpicture}
  =
  \frac{i\kappa_d^2}{4}
  \big\{p^{(+)}_\rho,V^{\mu_1\nu_1\mu_2\nu_2}(k_1,k_2)\big\}
  \ .
\end{align}
This gives rise to the usual rule of pulling out the overall momentum:
\begin{align}
  \big\{p^{(+)}_\rho,V^{\mu_1\nu_1\mu_2\nu_2}(k_1,k_2)\big\}
  =&
  -i
  (k_2-k_1)_\rho
  \\
  &\qquad\times
  \mathcal{V}^{\mu_1\nu_1\mu_2\nu_2}(k_1,-k_2)
  \ .
  \notag
\end{align}
Here, $(-)$ variables were put to zero on the right-hand-side with $\mathcal{V}^{\mu_1\nu_1\mu_2\nu_2}(k_1,-k_2)$ from eq.~\eqref{eq:curlyV} independent of $(-)$ variables.
The rule for
\begin{align}
  \big\{S^{(+)\alpha\beta},V^{\mu_1\nu_1\mu_2\nu_2}(k_1,k_2)\big\}
\end{align}
is somewhat more complicated but follows from $V^{\mu_1\nu_1\mu_2\nu_2}(k_1,k_2)$ and the brackets~\eqref{eq:PoissonBracketsPM}.
The vertex rule structure $V^{\mu_1\nu_1\mu_2\nu_2}(k_1,k_2)$ follows, as usual, from the action.

Carrying out these steps and using the known one-loop techniques, we find the results reported in the main text in eqs.~\eqref{eq:changeinmass} and~\eqref{eq:changeinspin}.

Note that our Poisson brackets are unconstrained and not the Dirac brackets which preserve constraints.
This implies that the absorptive tensorial observables do not necessarily preserve the covariant SSC.
We argue, however, that the change in the Pauli Lubanski vector,
\begin{align}
  \Delta a^\mu_{\rm abs}
  \, ,
\end{align}
is independent of the $Z_0^\mu$-dependence of the free energy $N_{\rm abs}$ with $Z_0^\mu$, the initial value of $Z^\mu(\tau)$.
This is the case, essentially, because the Poisson bracket of $a^\mu_0$ with $Z^\mu_0$ is proportional to $Z^\mu_0$.
Specifically, note that
\begin{align}
  \frac12 S_0^\mn S_{0\mn}
  =
  -m^2 a_0^\mu a_{0\mu} + Z_0^\mu Z_{0\mu}
  \, ,
\end{align}
and therefore that
\begin{align}
  \Delta(S^\mn S_\mn)
  =
  -4a_{0\mu} \Delta a^\mu_{\rm abs}
  +
  4Z_{0\mu} \Delta Z^\mu_{\rm abs}
  +
  \dots
  \, ,
\end{align}
with the dots signifying subleading terms in the PM expansion.
Imposing the covariant SSC on the initial state such that $Z_0^\mu=0$, the leading-order absorption of angular momentum computed in this way is independent of the value of $\Delta Z^\mu_{\rm abs}$.

\begin{widetext}
\section{Scattering phase shift and absorption factor}\label{app:BHPTResults}

In this appendix we present the explicit expressions for the elastic phase shift and absorption factor entering \cref{eq:BHPTScatteringInfo}, which make up the BHPT Magnusian \cref{eq:bhpt-expansion}.
These results were computed using the Nekrasov-Shatashvili construction of \cite{Bautista:2023sdf}.
All are accessible in machine-readable format with the supplementary material at \zenodo.

We organize the phase shift and absorption factor in an expansion in the PM parameter $\ePM$.
Additionally, up to order $\ePM^5$, these objects are polynomials in the spin parameter $\chi$. 

\subsection{Phase shift}
\label{app:phase-shift-expansion}

We write the double expansion of the phase shift in $\ePM$ and $\chi$ as
\begin{equation}
 {}_{-2} \delta_{\ell m}^P
  = \sum_{n=1}^{5}\ePM^n
    \sum_{k=0}^{n-1}\chi^k\,{}_{-2}\delta_{\ell m}^{P,(n,k)}
    +\mathcal O(\ePM^6),
  \label{eq:phase-shift-double-expansion}
\end{equation}
where logarithmic dependence on $\ePM$ is retained in ${}_{-2}\delta_{\ell m}^{P,(1,0)}$.
We remind the reader that we are not expanding in small $\chi$ here: the dependence of the phase shift is polynomial in $\chi$ up to this order, but $\cO(\ePM^6)$ corrections and beyond are not.
For compactness below, we introduce
\begin{equation}
  L\equiv\ell(\ell+1),
  \qquad
  \Psi_r(\ell)\equiv
  \psi^{(r)}(\ell-1)+\psi^{(r)}(\ell+3),
  \label{eq:phase-shift-notation}
\end{equation}
where $\psi^{(r)}$ is the polygamma function, with $\psi^{(0)}$ the
digamma function, and suppress the argument of $\Psi_r$ below.

The coefficients at first and second PM orders are
\begin{align}
  {}_{-2}\delta_{\ell m}^{P,(1,0)}
  &=\log(2\ePM)-\frac12-\frac{\Psi_0}{2}
    +\frac{3P}{L(L-2)},
    \label{eq:phase-d10}\\
  {}_{-2}\delta_{\ell m}^{P,(2,0)}
  &=\frac{\pi(24+13L+15L^2)}
    {4L(2\ell+1)(4L-3)},
    \label{eq:phase-d20}\\
  {}_{-2}\delta_{\ell m}^{P,(2,1)}
  &=-\frac{m}{2L}
    +\frac{6mP}{L^2(L-2)}.
    \label{eq:phase-d21}
\end{align}
At third order, we find
\begin{align}
  {}_{-2}\delta_{\ell m}^{P,(3,0)}
  &=\frac{\delta_{\ell m}^{(2,0)}}{\pi}\,\Psi_1
    +\frac{\Psi_2}{12}
    +\frac{3(2+3L)}{4L(4L-3)}
    -\frac{36P}{L^3(L-2)^3},
    \label{eq:phase-d30}\\
  {}_{-2}\delta_{\ell m}^{P,(3,1)}
  &=-\frac{\pi m(108+18L-L^2+5L^3)}
    {2(L-2)L^2(2\ell+1)(4L-3)},
    \label{eq:phase-d31}\\
  {}_{-2}\delta_{\ell m}^{P,(3,2)}
  &=\frac{-L^2(12+L)
      +(6+L)(2+3L)m^2}
    {4L^3(4L-3)}
    +\frac{3P\bigl[-(L-12)L^2
      +(-12-26L+11L^2)m^2\bigr]}
    {(L-2)L^4(4L-3)}.
    \label{eq:phase-d32}
\end{align}
The fourth-order coefficients read
\begin{align}
  \delta_{\ell m}^{P,(4,0)}
  &=\frac{\pi\mathcal P_{4,0}}
    {16(L-2)L^3(2\ell+1)^3
     (4L-15)(4L-3)^3},
    \label{eq:phase-d40}\\
  \delta_{\ell m}^{P,(4,1)}
  &=\frac{\delta_{\ell m}^{(3,1)}}{\pi}\,\Psi_1
    -\frac{m(81+16L+33L^2)}
      {6(L-2)L^2(4L-3)}
    -\frac{216mP}{L^4(L-2)^3},
    \label{eq:phase-d41}\\
  \delta_{\ell m}^{P,(4,2)}
  &=\frac{\pi\bigl[-L^2\mathcal P_{4,2}^{(0)}
       +3(L-2)\mathcal P_{4,2}^{(2)}m^2\bigr]}
    {8(L-2)L^4(2\ell+1)(4L-15)(4L-3)^2},
    \label{eq:phase-d42}\\
  \delta_{\ell m}^{P,(4,3)}
  &=\frac{m}{12(L-2)L^5(4L-3)}
    \Bigl[L^2(288+336L-L^2+3L^3)
    -(288+912L+432L^2+5L^4)m^2\Bigr]
    \notag\\
  &\quad+\frac{3mP}{L^6(L-2)^2(4L-3)}
    \Bigl[-L^2(96+160L-55L^2+3L^3)
    +(96+352L+176L^2-104L^3+13L^4)m^2\Bigr].
    \label{eq:phase-d43}
\end{align}
Finally, at fifth order,
\begin{align}
  \delta_{\ell m}^{P,(5,0)}
  &=\frac{\mathcal P_{5,0}}
      {48(L-2)L^3(4L-15)(4L-3)^3}
    +\frac{\delta_{\ell m}^{(4,0)}}{\pi}\,\Psi_1
    -\left(\frac{\delta_{\ell m}^{(2,0)}}{\pi}\right)^2\Psi_2
    -\frac{\delta_{\ell m}^{P,(2,0)}}{6\pi}\,\Psi_3
    -\frac{\Psi_4}{240}
    +\frac{3888P}{5L^5(L-2)^5},
    \label{eq:phase-d50}\\
  \delta_{\ell m}^{P,(5,1)}
  &=-\frac{3\pi m\mathcal P_{5,1}}
    {4L^4(L-2)^2(2\ell+1)^3
     (4L-15)(4L-3)^3},
    \label{eq:phase-d51}\\
  \delta_{\ell m}^{P,(5,2)}
  &=\frac{\delta_{\ell m}^{(4,2)}}{\pi}\,\Psi_1
    +\frac{L^2\mathcal P_{5,2}^{(0)}
        +3(L-2)\mathcal P_{5,2}^{(2)}m^2}
      {24(L-2)L^4(4L-15)(4L-3)^2}
    +\frac{108P\bigl[L^2(L-12)
        +(12+38L-27L^2)m^2\bigr]}
      {L^6(L-2)^3(4L-3)},
    \label{eq:phase-d52}\\
  \delta_{\ell m}^{P,(5,3)}
  &=\frac{\pi m\bigl[L^2\mathcal P_{5,3}^{(1)}
       +\mathcal P_{5,3}^{(3)}m^2\bigr]}
    {4(L-2)^2L^6(2\ell+1)(4L-15)(4L-3)^2},
    \label{eq:phase-d53}\\
  \delta_{\ell m}^{P,(5,4)}
  &=\frac{(L-2)L^4\mathcal P_{5,4}^{(0)}
       -2L^2\mathcal P_{5,4}^{(2)}m^2
       +\mathcal P_{5,4}^{(4)}m^4}
    {16(L-2)L^7(4L-15)(4L-3)^3}
    +\frac{3P\bigl[(L-12)(L-2)L^4
       \mathcal Q_{5,4}^{(0)}
       -2L^2\mathcal Q_{5,4}^{(2)}m^2
       +\mathcal Q_{5,4}^{(4)}m^4\bigr]}
    {4L^8(L-2)^2(4L-15)(4L-3)^3}.
    \label{eq:phase-d54}
\end{align}
At fourth and fifth orders we have introduced auxiliary polynomials which we define as
\begin{align}
  \mathcal P_{4,0}(L)
  &= -51840-102816L+953424L^2-51522L^3
    -123233L^4-85775L^5+2415L^6-61320L^7
    +18480L^8\\
  \mathcal P_{4,2}^{(0)}(L)
  &= -38880+7128L+1764L^2-1650L^3
     +218L^4+65L^5\\
  \mathcal P_{4,2}^{(2)}(L)
  &= 6480+19872L-2400L^2-436L^3
     +69L^4+85L^5\\
  \mathcal P_{5,0}(L)
  &= -38880+156168L+35640L^2-39114L^3
     +10743L^4-66775L^5+18372L^6\\
  \mathcal P_{5,1}(L)
  &= -77760-125064L+1450620L^2-650274L^3
    -145050L^4-42625L^5-22136L^6+18347L^7 \\
    &\quad-13944L^8+3696L^9\notag\\
  \mathcal P_{5,2}^{(0)}(L)
  &= 29160-4446L+9078L^2-1639L^3
    -573L^4\\
  \mathcal P_{5,2}^{(2)}(L)
  &= 4860+15054L+1855L^2+1226L^3
     +1025L^4\\
  \mathcal P_{5,3}^{(1)}(L)
  &= -155520-321408L+176508L^2-19434L^3
     -4991L^4+1324L^5+24L^6+51L^7\\
  \mathcal P_{5,3}^{(3)}(L)
  &= 155520+632448L+341568L^2-384288L^3
     +97068L^4-126L^5-2125L^6+80L^7
     -105L^8\\
  \mathcal P_{5,4}^{(0)}(L)
  &= -25920+32832L+40248L^2-5109L^3
    +547L^4+12L^5\\
  \mathcal P_{5,4}^{(2)}(L)
  &= 155520+36288L-666216L^2+188508L^3
     +200182L^4-23505L^5+1393L^6+100L^7\\
  \mathcal P_{5,4}^{(4)}(L)
  &= 259200+578880L-1188000L^2-1434480L^3
   +1217080L^4+337364L^5-45066L^6 \\
   &\quad+2763L^7+252L^8\notag\\
  \mathcal Q_{5,4}^{(0)}(L)
  &= -2160+1476L+5505L^2-1277L^3
     +44L^4\\
  \mathcal Q_{5,4}^{(2)}(L)
  &= -155520-114048L+798552L^2-84888L^3
     -522270L^4+188517L^5-21599L^6+708L^7\\
  \mathcal Q_{5,4}^{(4)}(L)
  &= -259200-708480L+1183680L^2+2172768L^3
    -1812832L^4-770680L^5+458808L^6 \\
    &\quad-69797L^7+3612L^8.\notag
\end{align}
Their $L$ arguments were suppressed in the phase shift coefficients above.
Subscripts correspond to the $\ePM$ and $\chi$ orders in which a polynomial appears;
superscripts distinguish the different powers of $m$ by which a polynomial is multiplied.

\subsection{Absorption factor}
\label{app:absorption-probability}

As the absorption probability is non-trivial from $\cO(\ePM^5)$, the result we need for this work is much more compact than the phase shift.
Up to this order, it receives contributions only from the $\ell=2$ harmonic:
\begin{equation}
  {}_{-2}\eta_{2,m}
  =1+\ePM^5\left[
      \chi\,\eta_m^{(5,1)}
      +\chi^3\eta_m^{(5,3)}
      +\chi^5\eta_m^{(5,5)}
    \right]+\mathcal O(\ePM^6).
  \label{eq:absorption-double-expansion}
\end{equation}
Its coefficients in this PM-spin expansion are
\begin{equation}
  \eta_m^{(5,1)}=\frac{m}{225},
  \qquad
  \eta_m^{(5,3)}=\frac{m(5m^2-8)}{900},
  \qquad
  \eta_m^{(5,5)}=\frac{m(m^2-1)(m^2-4)}{900}.
  \label{eq:absorption-coefficients}
\end{equation}
Notice that for physical values of $m$, $m=-2,-1,0,1,2$, $\eta_m^{(5,5)}$ completely vanishes.

\end{widetext}

\bibliographystyle{apsrev4-2}
\bibliography{FourLoopCompton}

\end{document}